\documentclass[twocolumn, aps,pre,superscriptaddress,showpacs,floatfix]{revtex4-2}

\UseRawInputEncoding
\usepackage{float}
\usepackage{dcolumn}
\usepackage{bm}
\usepackage[colorlinks = true, linkcolor = blue, urlcolor  = blue, citecolor = blue, anchorcolor = blue]{hyperref}
\usepackage{longtable}
\usepackage{booktabs}
\usepackage{mathrsfs}
\usepackage{graphicx,epsfig,latexsym,amssymb}
\usepackage{multirow,amsmath,array,booktabs,color}
\usepackage[section]{placeins}
\usepackage{color}
\usepackage{soul}

\begin{document}
\title{Dissecting Quantum Phase Transition in the Transverse Ising Model}
\author{Yun-Tong Yang}
\affiliation{School of Physical Science and Technology, Lanzhou University, Lanzhou 730000, China}
\affiliation{Lanzhou Center for Theoretical Physics, Key Laboratory of Theoretical Physics of Gansu Province, Key Laboratory of Quantum Theory and Applications of MoE, Gansu Provincial Research Center for Basic Disciplines of Quantum Physics, Lanzhou University, Lanzhou 730000, China}
\author{Fu-Zhou Chen}
\affiliation{School of Physical Science and Technology, Lanzhou University, Lanzhou 730000, China}
\affiliation{Lanzhou Center for Theoretical Physics, Key Laboratory of Theoretical Physics of Gansu Province, Key Laboratory of Quantum Theory and Applications of MoE, Gansu Provincial Research Center for Basic Disciplines of Quantum Physics, Lanzhou University, Lanzhou 730000, China}
\author{Hong-Gang Luo}
\email{luohg@lzu.edu.cn}
\affiliation{School of Physical Science and Technology, Lanzhou University, Lanzhou 730000, China}
\affiliation{Lanzhou Center for Theoretical Physics, Key Laboratory of Theoretical Physics of Gansu Province, Key Laboratory of Quantum Theory and Applications of MoE, Gansu Provincial Research Center for Basic Disciplines of Quantum Physics, Lanzhou University, Lanzhou 730000, China}

\begin{abstract}
Despite the fact that a complete theoretical description of critical phenomena in connection with phase transitions has been well-established through the renormalization group theory, the microscopic nature of the phase transitions remains to be understood in a satisfactory way. For example, how does the interaction between individuals drive a system from one phase to another as a specific parameter varies, and how do the individuals respond to changes in this parameter during the process? Here we take the well-studied quantum phase transition (QPT) in the one-dimensional transverse Ising model (TIM) as an example to exhibit such a microscopic process. We first introduce $2L$ collective structures,referred to as patterns, for the TIM with $L$ ferromagnetically interacting spins, and then analyze the contributions of these patterns to the system's states, e.g., the ground state, the first excited state, and so on, from which the analogue of the QPT process between the disordered phase in the weakly coupling regime and the ferromagnetic phase in the strongly coupling regime is clearly identified around the interaction strength $J_c =1$. We systematically explore this process for small lattice sizes of $L=6, 8, 10, 12$, whose ground state energies are identical to those obtained by direct numerical exact diagonalization. Increasing the system size up to $L=128$, the actual QPT point located at $J_c = 1$ in the thermodynamical limit is gradually approached. Our results show that the pattern picture is not only able to provide a microscopic process of phase transitions, but also of practical interest in analyzing analogues of QPT in diverse quantum simulation platforms.
\end{abstract}

\pacs{}
\maketitle

\section{\label{sec:level1}Introduction}
The investigations of phase and phase transition of matter in nature or laboratory are one of central issues in modern physics, especially in condensed matter physics, statistical physics as well as complex systems by various ways ranging from theoretical models, experimental measurements, numerical methods to quantum simulations. Conceptually, the order parameter and associated symmetry breaking are the foundation to understand and describe phenomenologically the phase transitions \cite{Landau1937, Landau1980}, and methodologically the renormalization group built by Wilson in 1970's \cite{Wilson1974} provides a complete theoretical description of the critical phenomena to occur in the vicinity of the phase transition points. 

The most important properties of critical phenomena are universality and scaling behaviors, which classify phase transitions based on critical exponents \cite{Chaikin2000}, independent of the specific details of the system's Hamiltonian. Thus it is commonly believed that the critical phenomena in connection with phase transitions have been sufficiently described. However, the understanding of the phase transition itself, both thermodynamically and quantum mechanically, remains incomplete \cite{Kastner2008}. While most of the existing literature focus on the classification of different types of phase transition in various physical systems, few works put emphasis on the mechanism or nature of phase transition or related issues. It should be pointed out that Lee and Yang \cite{Yang1952} developed a theory of equations of state and thermodynamic phase transitions, which relates the zeros of the grand canonical partition function in the complex fugacity plane to nonanalyticity of the corresponding thermodynamic function. However, sufficient or necessary conditions for the occurrence of a phase transition, to a large extent, remains to be explored. For more details, one can refer to the review in Ref.  \cite{Kastner2008}. 

On the other hand, the dynamical process of phase transitions, known as the Kibble-Zurek mechanism, has been proposed by Kibble and Zurek \cite{Kibble1976, Zurek1985, Zurek1996}. This mechanism identifies two key features characterizing the phase transition process: critical slowing down and the growth of the correlation length. The properties of the post-transition broken symmetry state can be inferred from such a dynamical process. Although the Kibble-Zurek mechanism has been widely tested in various systems \cite{Weiler2008, Pyka2013, Deutschlander2015, Keesling2019, Ko2019, Schmitt2022}, both experimentally and numerically, its validity and/or applied ranges still remain to be clarified \cite{Braun2015}. Furthermore, fundamental questions regarding the microscopic process of phase transitions$-$such as why macroscopic numbers of particles \textit{collectively and simultaneously} transform from one phase to another, and how the new phase emerges from the old one$-$still lack satisfactory explanations. 

These questions highlight the need for a deeper exploration into the nature of phase transitions and the mechanisms driving them. Motivated by this, we propose the pattern picture to explore the microscopic process of phase transitions. We have previously applied the pattern picture to the quantum Rabi model to study the superradiant phase transition \cite{Yang2022b}. In this work, we extend this approach to a prototypical model, namely, the one-dimensional transverse-field Ising model (TIM), which is not only of theoretical interest \cite{Sachdev2011, Dutta2015}, but also has been realized in realistic materials \cite{Coldea2010, Breunig2017}. More importantly, some well-known or unknown physics involved in such a simple model can be simulated or unveiled by various quantum spin simulators \cite{Friedenauer2008, Islam2011, Kim2011, Johnson2011, Georgescu2014, Kandala2017, Monroe2021}. It is well-known that the TIM is integrable and has been exactly solved by employing the Jordan-Winger transformation \cite{Pfeuty1970}. The result shows that this model exhibits a quantum phase transition (QPT) from the paramagnetic (disordered) phase at weak Ising interacting strength to the ferromagnetic (ordered) phase at strong interaction for a given transverse field, even for small simulating systems. For example, onset of a QPT has been observed by using nine trapped ions quantum simulator \cite{Islam2011, Kim2011}, and the ground state of an artificial Ising spin system comprising an array of eight superconducting flux quantum bits has been obtained by using quantum annealing \cite{Johnson2011}. Although this QPT has been thoroughly studied from a theoretical perspective, our aim here is to dissect this phase transition and elucidate its microscopic process by employing the pattern picture.  

The method we adopt consists of two successive diagonalizations. For the TIM with $L$ interacting spins, the first diagonalization is performed in an operator space to obtain fundamental patterns, characterized by the patterns' eigenvalues $\lambda_n$ and the ordered spin operators $\{i\hat\sigma^y_1, \hat\sigma^z_1, i\hat\sigma^y_2, \hat\sigma^z_2, \cdots, i\hat\sigma^y_L, \hat\sigma^z_L\}$. The obtained new Hamiltonian is proved to be completely equivalent to the original one. The second diagonalization is performed in the usual spin basis, as done in direct numerical exact diagonalization. With the patterns at hand, we analyze the microscopic process of analogue of phase transition and identify the roles played by different patterns as varying the Ising interacting strength $J$. This is what means in the title that we dissect the QPT of the TIM. In addition, we present the pattern occupancy to further show how the analogue of phase transition occurs underlying the pattern picture. It is straightforward to extend the pattern picture to higher dimensional cases and other many-body models in order to further explore the intriguing physics of phase transitions involved in those models. 

The paper is organized as follows. In Sec. \ref{sec:level2}, we introduce the TIM and outline the method used to derive patterns. In Sec. \ref{sec:level3}, we present patterns' information for finite lattice sizes of $L=6, 8, 10, 12$ and analyze their properties. In Sec. \ref{sec:level4}, we investigate the analogue of ground-state phase transition for these finite lattice sizes using the pattern picture. In Sec. \ref{sec:level5}, we extend our analysis to larger lattice sizes of $L=16, 32, 64$, and $128$ in order to exhibit how the analogue of the QPT studied above approaches to the real QPT in the thermodynamic limit. Finally, Sec. \ref{Conclusion} is devoted to conclusion and discussion.

\section{\label{sec:level2}Model and Method}
The Hamiltonian of the TIM reads
\begin{equation}
\hat{H}' = - J'\sum_{i,\delta}\hat\sigma^z_i \hat\sigma^z_{i+\delta} - g\sum_i \hat\sigma^x_i, \label{Ising0}
\end{equation}
where $J'$ and $g$ are the Ising interaction between two spins denoting by Pauli matrix $\hat\sigma$ located at site $i$ and its nearest neighbors $i+\delta$ and transverse fields, respectively. They are assumed to be greater than or equal to zero, and in this case the Ising interaction is ferromagnetic. For convenience, we reformulate Eq. (\ref{Ising0}) as $\hat{H}'= \frac{g}{2} \hat{H}$, which means that in the following formalism we take $g/2$ as the units of energy, and thus one has 
\begin{equation}
\hat{H} = - J\sum_{i,\delta} \left(\hat\sigma^z_i \hat\sigma^z_{i+\delta} + \hat\sigma^z_{i+\delta} \hat\sigma^z_i\right) -2 \sum_{i}\hat{\sigma}^x_i,\label{Ising1}
\end{equation}
where $J = J'/g$. To more clearly illustrate the physical picture of pattern formulation, here we limit ourselves to the one-dimensional case and $\delta=1$. For the chain length $L$ with periodic boundary condition, the model Hamiltonian can be reformulated as
\begin{eqnarray}
\hat{H} &=& \left(
\begin{array}{ccccccc}
 -i\hat{\sigma}^y_1& \hat{\sigma}^z_1& -i\hat{\sigma}^y_2& \hat{\sigma}^z_2& \cdots & -i\hat{\sigma}^y_L& \hat{\sigma}^z_L
\end{array}
\right)\nonumber\\
&\times&
\left(
\begin{array}{ccccccc}
0 &- 1 &0 &0 &\cdots &0 &0 \\
- 1 &0 &0 &- J &\cdots &0 &- J \\
0 &0 &0 &- 1 &\cdots &0 &0 \\
0 &- J &- 1 &0 &\cdots &0 &0 \\
\vdots &\vdots &\vdots &\vdots &\ddots &\vdots &\vdots \\
0 &0 &0 &0 &\cdots &0 &- 1 \\
0 &- J &0 &0 &\cdots &- 1 &0 
\end{array}
\right)
\left(
\begin{array}{c}
i\hat{\sigma}^y_1\\
\hat{\sigma}^z_1\\
i\hat{\sigma}^y_2\\
\hat{\sigma}^z_2\\
\vdots\\
i\hat{\sigma}^y_L\\
\hat{\sigma}^z_L
\end{array}
\right),\label{Ising2}
\end{eqnarray}
where the identity of Pauli matrices $\hat{\sigma}^y \hat{\sigma}^z = i \hat{\sigma}^x$ has been used for each site $i$. Here we point out that the letter $i$ preceding a Pauli operator denotes the imaginary unit, while as a subscript of a Pauli operator, it denotes the lattice site. The matrix in Eq. (\ref{Ising2}) has dimension $2L \times 2L$, and it can be diagonalized to obtain eigenvalues and corresponding eigenfunctions $\{\lambda_n, u_n\} (n = 1, 2, \cdots, 2L)$. Thus the Hamiltonian \eqref{Ising1} can be rewritten as
\begin{equation}
\hat{H} = \sum_{n=1}^{2L} \lambda_n \hat{A}^\dagger_n \hat{A}_n, \label{Ising3a} 
\end{equation}
where the obtained $\hat{A}_n$ composes of single-body operators
\begin{equation}
\hat{A}_n = \sum_{i=1}^{L}\left[u_{n,2i-1} (i\hat{\sigma}^y_{i}) + u_{n,2i} \hat{\sigma}^z_{i}\right]. \label{Ising3b}
\end{equation}
We call $\hat{A}_n$ patterns in the following marked by $\lambda_n$. Obviously, there are two spin components $(i\hat\sigma^y_i, \hat\sigma^z_i)$ for each lattice site $i$, where $\hat\sigma^z_i$ represents spin-up/down and $i\hat\sigma^y_i$ represents the flip of spins.

\section{\label{sec:level3}Patterns' Information} 
\begin{figure}[tbp]
\begin{center}
\includegraphics[width = 0.9\columnwidth]{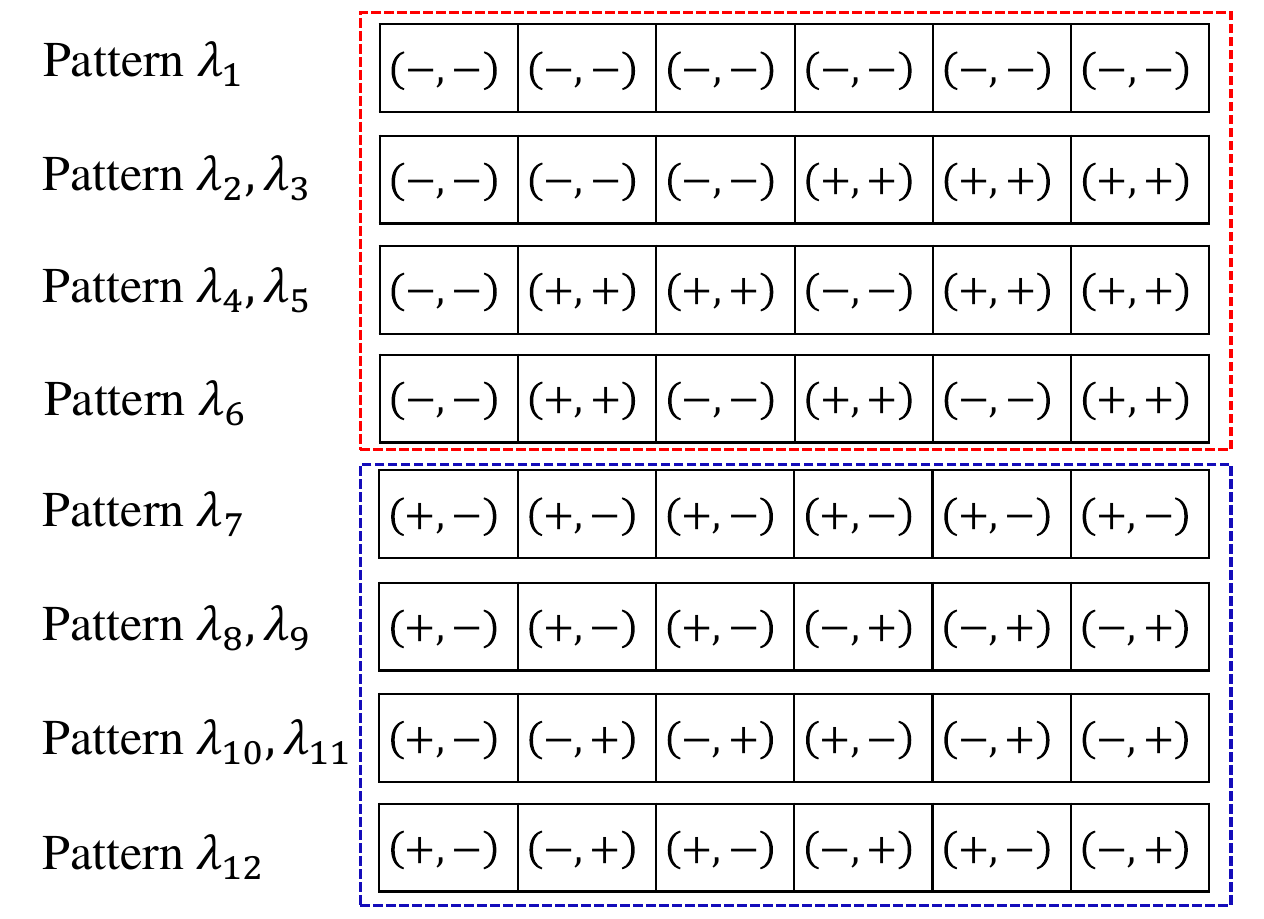}
\caption{The patterns and their relative phases for $L=6$ obtained by the first diagonalization, marked by the single-body operators $\hat{A}_n = \sum_{i=1}^L \left[u_{n,2i-1} (i\hat{\sigma}^y_i) + u_{n,2i}\hat{\sigma}^z_i\right]$ with $(\pm,\pm)$ denoting the signs of $(u_{n,2i-1},u_{n,2i})$. All patterns are divided into two groups marked by the red and blue frames with $\lambda_n < 0$ and $\lambda_n > 0$, respectively. The characteristic difference between these two groups is that for the pattern with $\lambda_n < 0$ (red frame) the operators within the sites are in-phase, but for the pattern with $\lambda_n > 0$ (blue frame) they are out-of-phase. For these two groups of patterns, different patterns are distinguished by the phases between site $i$ and its nearest neighbor sites, which form domains if they are in-phase for $\hat{\sigma}^z_i$, otherwise, kinks if they are out-of-phase for $\hat{\sigma}^z_i$. Here it should be mentioned that the eigenvectors $(u_{n,2i-1},u_{n,2i})$ are free of a total phase factor $e^{i\pi}$ but their relative phases remain fixed.}\label{fig1}
\end{center}
\end{figure}

\begin{figure}[tbp]
\begin{center}
\includegraphics[width = \columnwidth]{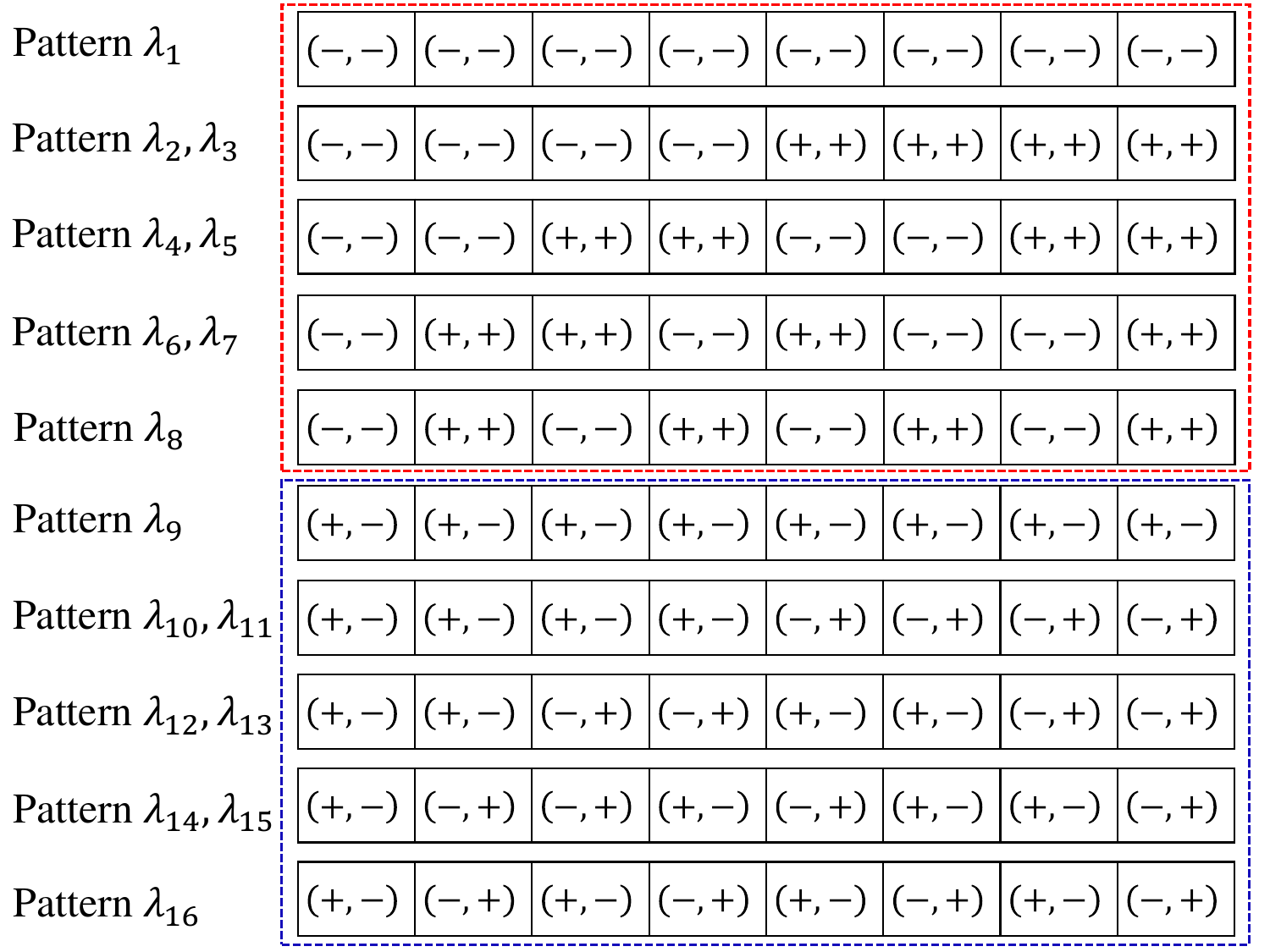}
\caption{The patterns and their relative phases for $L=8$. Other information is consistent with Fig. \ref{fig1}.}\label{fig2}
\end{center}
\end{figure}

\begin{figure}
\begin{center}
\includegraphics[width = \columnwidth]{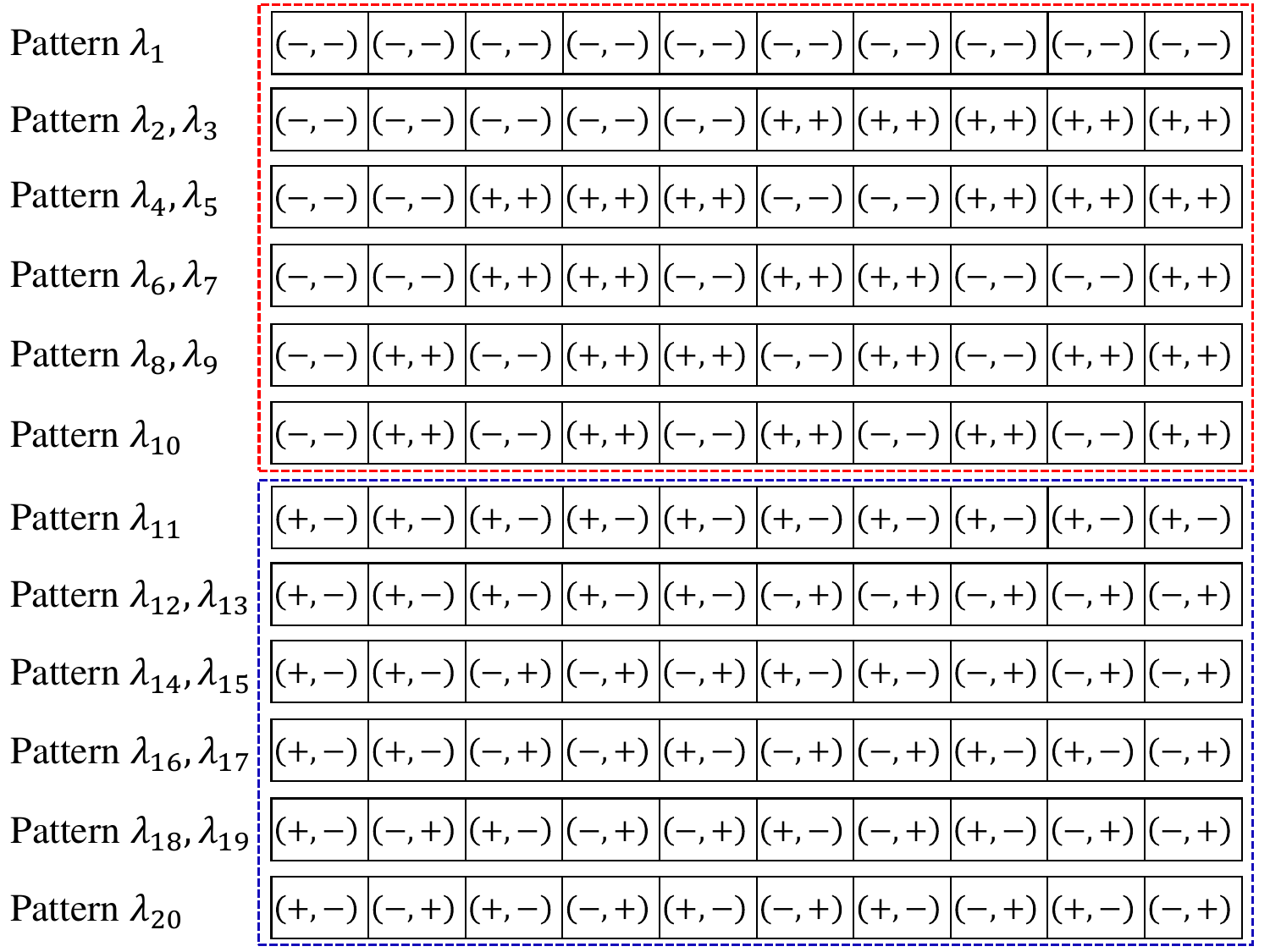}
\caption{The patterns and their relative phases for $L=10$. Other information is consistent with Fig. \ref{fig1}.}\label{fig3}
\end{center}
\end{figure}

\begin{figure}
\begin{center}
\includegraphics[width = \columnwidth]{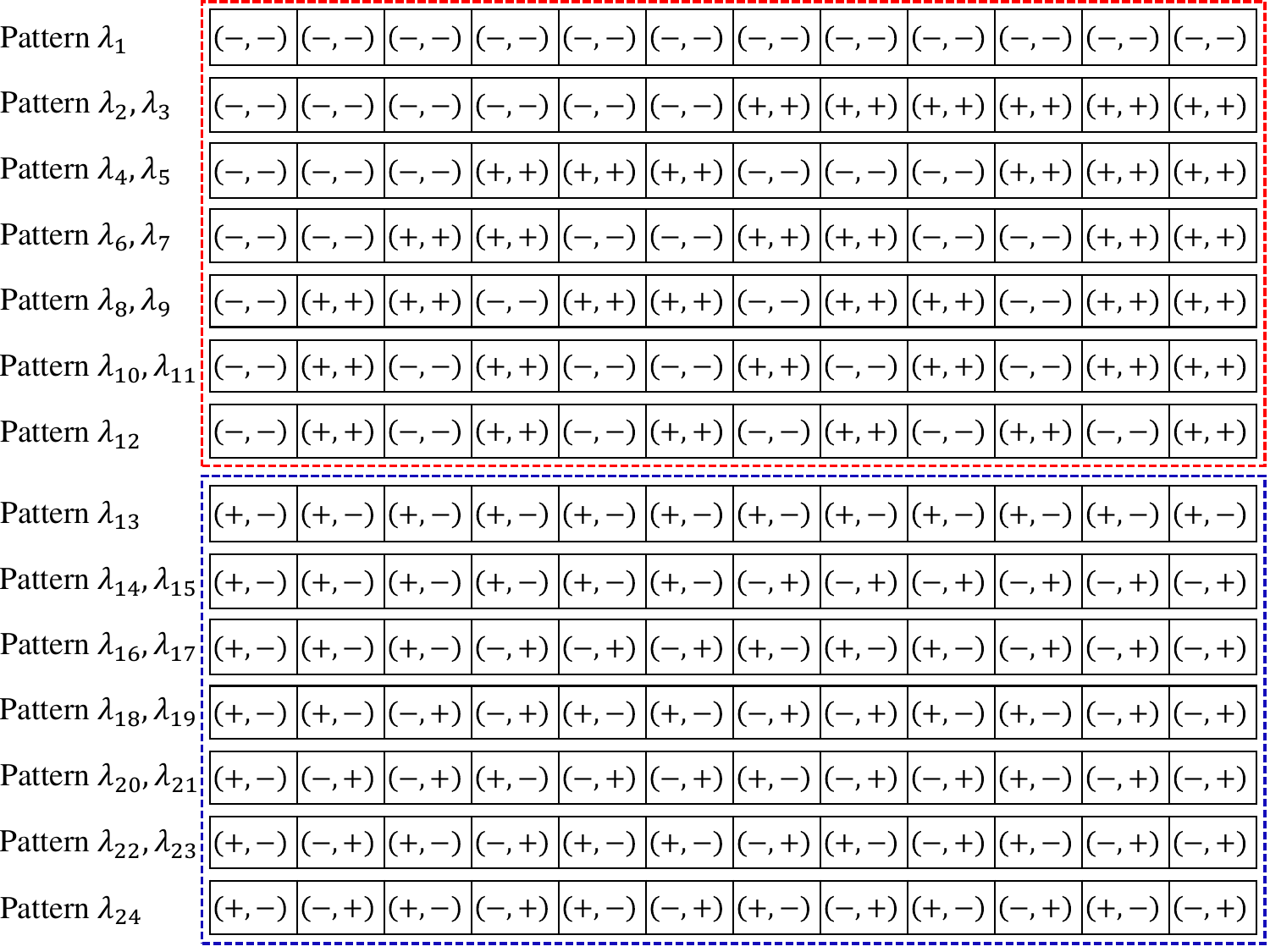}
\caption{The patterns and their relative phases for $L=12$. Other information is consistent with Fig. \ref{fig1}.}\label{fig4}
\end{center}
\end{figure}
 
\begin{figure}[tbp]
\begin{center}
\includegraphics[width = \columnwidth]{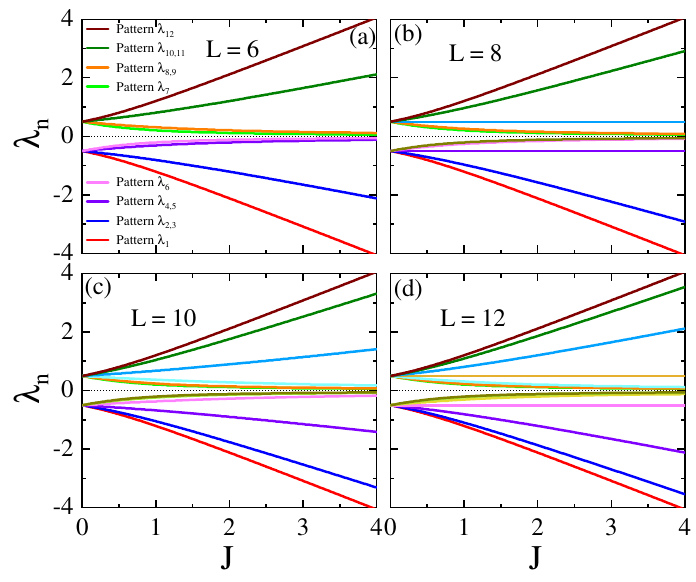}
\caption{The eigenenergies $\lambda_n$ of the patterns as functions of the Ising interacting strength $J$ for $L =6, 8, 10, 12$. The patterns are divided into two groups, which have positive and negative eigenenergies, respectively, satisfying with $\lambda_n = - \lambda_{2L - n +1}$ ($n = 1, 2, \cdots, 2L$). There are still degenerate for the patterns such as $\lambda_{2,3}$, $\lambda_{4,5}$, $\cdots$, and their positive eigenenergy counterparts. For simplicity, we only give the legend of (a), from bottom to top (or from the red line to wine line) corresponding pattern $\lambda_1$, $\lambda_{2,3}$, $\lambda_{4,5}$, $\lambda_6$, $\lambda_{7}$, $\lambda_{8,9}$, $\lambda_{10,11}$, $\lambda_{12}$. The legend of (b): from bottom to top corresponding pattern $\lambda_1$, $\lambda_{2,3}$, $\lambda_{4,5}$, $\lambda_{6,7}$, $\lambda_8$, $\lambda_{9}$, $\lambda_{10,11}$, $\lambda_{12,13}$, $\lambda_{14,15}$, $\lambda_{16}$. The legend of (c): from bottom to top corresponding pattern $\lambda_1$, $\lambda_{2,3}$, $\lambda_{4,5}$, $\lambda_{6,7}$, $\lambda_{8,9}$, $\lambda_{10}$, $\lambda_{11}$, $\lambda_{12,13}$, $\lambda_{14,15}$, $\lambda_{16,17}$, $\lambda_{18,19}$, $\lambda_{20}$. The legend of (d): from bottom to top corresponding pattern $\lambda_1$, $\lambda_{2,3}$, $\lambda_{4,5}$, $\lambda_{6,7}$, $\lambda_{8,9}$, $\lambda_{10,11}$, $\lambda_{12}$, $\lambda_{13}$, $\lambda_{14,15}$, $\lambda_{16,17}$, $\lambda_{18,19}$, $\lambda_{20,21}$, $\lambda_{22,23}$, $\lambda_{24}$.}\label{fig5}
\end{center}
\end{figure}

Firstly we explore the properties of the patterns by extracting signs (i.e. plus or minus) of eigenvectors $(u_{n,2i-1},u_{n,2i})$ to characterize the relationship between lattice spins and their change from one pattern to another with different eigenvalues. In order to show patterns' evolution with lattice sizes, we choose four different lattice sizes of $L=6, 8, 10$, and $12$, as shown in Fig. \ref{fig1} to Fig. \ref{fig4}. The patterns' eigenvalue $\lambda_n$ varying with the interaction strength $J$ is shown in Fig. \ref{fig5}. According to the signs of the eigenvalues $\lambda_n$, all patterns can be divided into two groups, one has negative eigenvalues $\lambda_n < 0$, as marked by red dashed frame in Fig. \ref{fig1} to Fig. \ref{fig4}, and the other has positive eigenvalues $\lambda_n > 0$ marked by blue dashed frame. For a lattice size of $L$, there are $L$ pairs of signs (plus and minus) for each pattern $\lambda_n$, denoting the relative phase of the eigenvectors $u_{n,2i-1}$ and $u_{n,2i}$. It is quite interesting to see a few characteristic features for these relative phases in the patterns obtained: (i) for all patterns with $\lambda_n < 0$, the $u_{n,2i-1}$ and $u_{n,2i}$ have the same signs, namely, the eigenvectors for each site are in-phase; on the contrary, they have opposite signs for the patterns $\lambda_n > 0$, that is to say, they are out-of-phase; (ii) for the pattern $\lambda_n < 0$, the pattern $\lambda_1$ has only one-domain, namely, all sites have the same phase, a typical character of a ferromagnetic order. It dominates over all other patterns in the ferromagnetic phase (or strong interaction regime in our presentation); (iii) from the pattern $\lambda_{2,3}$ to $\lambda_L$, the number of domains reads $2$, $4$, $6$, $\cdots$, $L$, which corresponds successively to $2$, $4$, $6$, $\cdots$, $L$ kinks, with possible different orders of the kinks in degenerate cases such as the patterns $\lambda_{2,3}$, $\lambda_{4,5}$, and so on; (iv) in the pattern $\lambda_L$, the phase is opposite for each pairs of nearest neighbor sites, which corresponds to the opposite limit of the pattern $\lambda_1$; (v) similar behaviors in the group of patterns $\lambda_n > 0$ have also been observed in the lower blue frame.  

Besides the above observations, it is important to emphasize that the number of patterns is linearly proportional to the system size, making it feasible to obtain all patterns even for larger systems. While the total number of patterns increases as the system size grows, certain fundamental characteristics of the patterns remain independent of the system size. For instance, the pattern $\lambda_1$ always exhibits a single-domain structure, the patterns $\lambda_{2,3}$ retain a two-domain structure, and so on. Additionally, the ordering of patterns based on their eigenenergies$-$where the pattern $\lambda_1$ has the lowest eigenenergy, followed by the pattern $\lambda_{2,3}$, the pattern $\lambda_{4,5}$, and so on up to the pattern $\lambda_{2L}$$-$is also unaffected by the system size, as shown in Fig. \ref{fig5}. These robust features of the patterns make the pattern picture a highly convenient framework for describing the analogue of phase transitions, as further elaborated in the following section.

\section{\label{sec:level4} Analogue of Phase Transition}
After clarifying the pattern picture obtained, we solve Eq. (\ref{Ising3a}) by inserting into the complete basis $|\{\sigma^z_i\}\rangle (i = 1, 2,\cdots, L)$ with $\hat{\sigma}^z_i |\{\sigma^z_i\}\rangle = \pm_i(\uparrow,\downarrow) |\{\sigma^z_i\}\rangle$. Firstly, one readily obtains the matrix $\left[\hat{A}_n\right]_{\{\sigma^z_i\},\{\sigma^z_i\}^{\prime}} = \langle\{\sigma^z_i\}|\hat{A}_n|\{\sigma^z_i\}^{\prime}\rangle$; then Eq. (\ref{Ising3a}) can be solved by diagonalizing the matrix obtained by
\begin{eqnarray}
&& \langle \{\sigma^z_i\}|\hat{H}|\{\sigma^z_i\}^{\prime}\rangle = \langle \{\sigma^z_i\}|\sum_{n=1}^{2L} \lambda_n \hat{A}^\dagger_n \hat{A}_n|\{\sigma^z_i\}^{\prime}\rangle \nonumber\\
&& \hspace{0.cm} = \sum_{n=1}^{2L} \lambda_n \langle \{\sigma^z_i\}|\hat{A}^\dagger_n \hat{A}_n|\{\sigma^z_i\}^{\prime}\rangle \nonumber\\
&& \hspace{0.cm} = \sum_{n=1}^{2L} \lambda_n  \sum_{\{\sigma^z_i\}^{\prime\prime}} \left[\hat{A}^\dagger_n\right]_{\{\sigma^z_i\},\{\sigma^z_i\}^{\prime\prime}}\left[\hat{A}_n\right]_{\{\sigma^z_i\}^{\prime\prime},\{\sigma^z_i\}^{\prime}}.\label{Ising4}
\end{eqnarray}
After obtaining the eigen-wavefunctions $\Psi_i$ ($i = 0, 1, \cdots$, corresponding to the ground state, the first excited state, and so on, respectively), what we do is to project the wavefunction onto different patterns in order to calculate the contributions of different patterns to the interested physical quantities. For example, the energy contributions of different patterns to the ground state $\Psi_0$ are calculated by the formula
\begin{equation}
E_{\lambda_n} = \lambda_n \langle \Psi_0| \hat{A}^\dagger_n \hat{A}_n|\Psi_0 \rangle, (n = 1,2,\cdots,2L). \label{Ei}
\end{equation}

Figure \ref{fig6} (a1), (b1), (c1) and (d1) present the results for the ground state as functions of the Ising interaction $J$, respectively, for $L = 6, 8, 10, 12$, as shown by heavy black solid lines. In order to confirm the validity of the pattern formulation, the results of direct exact diagonalization have also presented for comparison, shown as circles. The exact agreement between them from weak to strong interaction regimes is noticed, which is not surprising since no any approximation has been introduced. 

\subsection{\label{sec:level4A} Pattern components of the ground state energy}
\begin{figure*}[tbp]
\begin{center}
\includegraphics[width = 1.8\columnwidth]{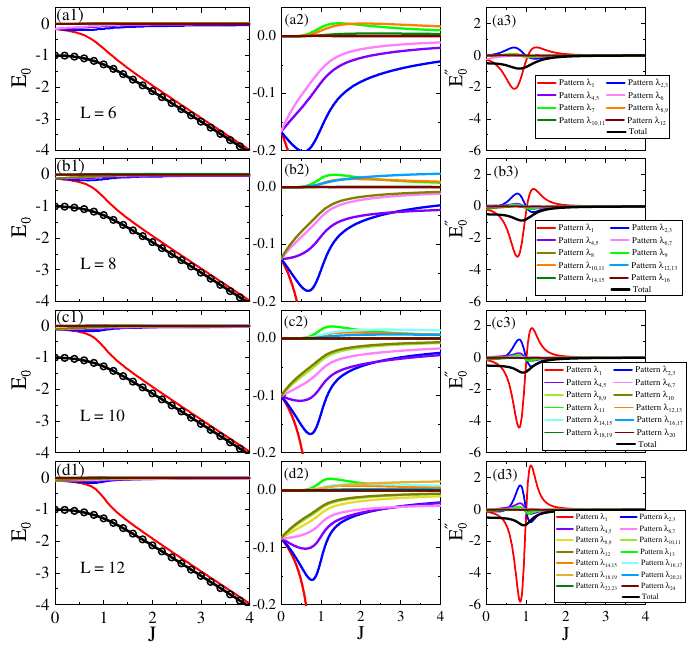}
\caption{The first column[(a1), (b1), (c1), (d1)]: The ground state energy (black solid lines) and the corresponding pattern components (colored solid lines) for different system sizes of $L = 6, 8, 10$, and $12$. The second column[(a2), (b2), (c2), (d2)]: The corresponding enlarged plots of the first column. The third column[(a3), (b3), (c3), (d3)]: The second-order derivatives of the ground state energy and their pattern components. Circles represent the results of numerical exact diagonalization. }\label{fig6}
\end{center}
\end{figure*}

It is well-known that the one-dimensional TIM exhibits a QPT from ferromagnetic phase to paramagnetic one at critical transverse field $g_c$ \cite{Sachdev2011} for the ground state at zero temperature. Due to the energy unit we used here the QPT occurs around $J = J_c =1$. For the analogue of QPT presented here, it occurs as the first excited state and the ground state become almost degenerate. In the weak interaction regime, the system is in the paramagnetic regime and it goes cross to the ferromagnetic phase at the strong interaction regime as $J > J_c$. It is more interesting to check this analogue of phase transition by the pattern picture. 

Figure \ref{fig6} presents the ground state energy and the corresponding pattern components, calculated using Eq. (\ref{Ei}), for different lattice sizes of $L=6, 8, 10$, and $12$. The figure is organized into three columns: the first column [(a1), (b1), (c1) and (d1)] provides an overall view of the results; the second column [(a2), (b2), (c2) and (d2)] displays the corresponding enlarged plots; the third column [(a3), (b3), (c3) and (d3)] shows the second derivatives of the ground state energy and the patterns' components as a function of Ising interaction $J$. It is noted that: (i) At $J=0$, the patterns in the same group are degenerate, in which the transverse field term determines only the phases in each site. The in-phase of the $i\hat{\sigma}^y_i$ and $\hat{\sigma}^z_i$ has a lower energy and the out-of-phase of them has a higher energy. (ii) With increasing $J$, the degeneracies are lifted. In the group of lower energy the pattern $\lambda_1$ separates itself from the other patterns by rapidly decreasing its energy, shown as red solid lines in Fig. \ref{fig6} (a1), (b1), (c1) and (d1). The energy contributions of the other patterns in this group become more and more less, even the patterns $\lambda_{2,3}$ and $\lambda_{4,5}$ have a slight increasing at the beginning, but turn back once $J$ approaches to its critical point, as shown in (a2), (b2), (c2) and (d2). (iii) In the group of higher energy the pattern $\lambda_{L+1}$ marked by the green line has an apparent response to the change of the interacting strength, and the responses of other patterns in this group are unimportant, especially for the larger lattice sizes such as $L=12$. (iv) As $J$ goes across the critical point, the energy contributions of all patterns except for the pattern $\lambda_1$ become more and more less, and thus the pattern $\lambda_1$ dominates. It is not surprising since the pattern $\lambda_1$ has a character of long-range ferromagnetic order, which corresponds to the ferromagnetic ground state of the system. This analogue of phase transition is more clearly seen by the second derivative of the ground state energy shown in Fig. \ref{fig6} (a3), (b3), (c3) and (d3) as black solid lines.

From the above observations, one can clearly see how the long-range order builds in the process of phase transition. In the small $J$ regime, all patterns in the $\lambda_n<0$ group exist. These patterns have different domain structures, thus the direction of spins is random. As a result, the system is in the paramagnetic phase. As $J$ increases, the energy contributions of certain patterns such as the pattern $\lambda_{2,3}$ for $L = 6, 8$ and the newly added patterns $\lambda_{4,5}$ for $L = 10, 12$ increase at the beginning and then decrease. Other patterns decrease in a monotonic way, in which the decrease of patterns with less domains is not so rapid in comparison to that of patterns with more domains. This indicates that the patterns with more domains are removed from the system first and then those with fewer domains are removed. After finishing the analogue of phase transition, the system is mainly left with the pattern $\lambda_1$ which has a single-domain structure. Naturally, the system is in the analogue of ferromagnetic phase. In addition, it is interesting to check the behavior of the pattern $\lambda_{L+1}$ in the $\lambda_n>0$ group, which has a character of in-phase of inter-sites, but in the intra-sites the two operators of $i\hat{\sigma}^y_i$ and $\hat{\sigma}^z_i$ are out-of-phase. It is noticed that after the phases of inter-sites have been built up, namely, the analogue of long-range order forms, the in-phase in each site just begins to form gradually and it lasts in a large interacting strength range, as shown in Fig. \ref{fig6} (a2), (b2), (c2) and (d2) marked by the green line. This is exactly the process of building the long-range order during the phase transition. It should point out that the pattern $\lambda_{L+1}$ plays the role of quantum fluctuations, which is uncovered quantitatively for the first time by identifying explicitly the corresponding pattern. 

Based on the above discussion, we now elaborate on the meaning of the term ``dissecting'' in the title of the paper. Conventionally, detecting a QPT in a system involves calculating the ground state energy or other relevant physical quantities and analyzing their derivatives to identify singularities. In contrast, the pattern picture offers a fundamentally different approach. It decomposes the system's Hamiltonian into distinct components, each of which exhibits a markedly different response to changes of the system's parameter. For example, in the TIM we study here, the pattern $\lambda_1$ rapidly acquires negative energy, driving the system toward a lower ground state energy. Conversely, the pattern $\lambda_{L+1}$ also responds rapidly but contributes positive energy, thereby competing with the pattern $\lambda_1$. In comparison to these two patterns, the other patterns play a relatively minor role in the ground state of the system. By dissecting the Hamiltonian in this manner, we can directly observe how each pattern contributes to the ground state and how these contributions evolve with the coupling strength $J$. The behaviors of these patterns allow us to determine the occurrence of a phase transition, especially in larger lattice sizes, which will be discussed in the next section. This is exactly the meaning of the word ``dissecting'' in our work, and obviously, this approach differs from the conventional approach in detecting and describing the phase transition. In a word, the pattern picture provides a direct and intuitive way to reveal the microscopic process underlying QPTs, at least their analogues. 

\subsection{\label{sec:level4B} Patterns' occupancy of the ground state}
Besides the ground state energy and the corresponding pattern components, the phase transition can also be observed from the patterns' occupancy of the ground state, which is defined as
\begin{equation}
\langle\hat O_0\rangle = \langle\Psi_0|\hat{A}^\dagger_n \hat{A}_n|\Psi_0\rangle, (n = 1,2,\cdots,2L),\label{Occ}
\end{equation}
where $|\Psi_0\rangle$ is the ground state eigen-wavefunction of the system. 

\begin{figure}[tbp]
\begin{center}
\includegraphics[width = \columnwidth]{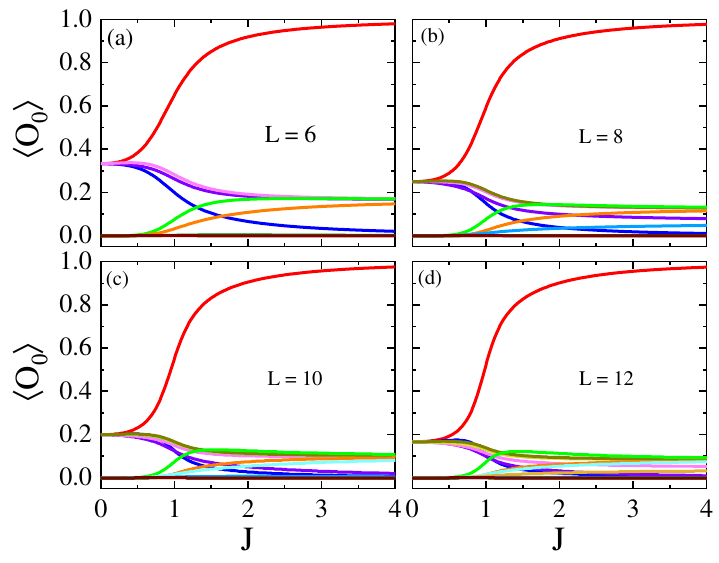}
\caption{The patterns' occupancy in the ground state of the system with (a) $L=6$, (b) $L=8$, (c) $L=10$ and (d) $L=12$. The legend is consistent with Fig. \ref{fig6}. The QPT occurs around $J = 1.0$.}\label{fig7}
\end{center}
\end{figure}

Figure \ref{fig7} shows the patterns' occupancy of the ground state for lattice sizes of $L=6, 8, 10$, and $12$, which behave similarly to the pattern components of energy, as depicted in Fig. \ref{fig6}. At $J=0$, patterns within the same group are degenerate. For small value of $J$, all patterns in the $\lambda_n < 0$ group contribute equally to the system, while patterns in the $\lambda_n > 0$ group make no contribution. Near the phase transition point at $J=1.0$, the pattern $\lambda_1$ becomes increasingly significant and begins to separates from the other patterns, as indicated by the red lines in Fig. \ref{fig7}. Beyond the phase transition, the pattern $\lambda_1$ dominates over other patterns. Notably, the pattern $\lambda_{L+1}$ exhibits a pronounced increase near the phase transition point, followed by a gradual decrease. This behavior reflects the role of quantum fluctuations, as discussed in the preceding section. Thus, the patterns' occupancy serves as a valuable physical quantity for elucidating the microscopic process of the QPT in the TIM. This interesting observation should be experimentally tested in quantum spin simulators such as the ion-trap \cite{Monroe2021}.

\subsection{\label{sec:level4C} Pattern components of excited states}
\begin{figure}[tbp]
\begin{center}
\includegraphics[width = \columnwidth]{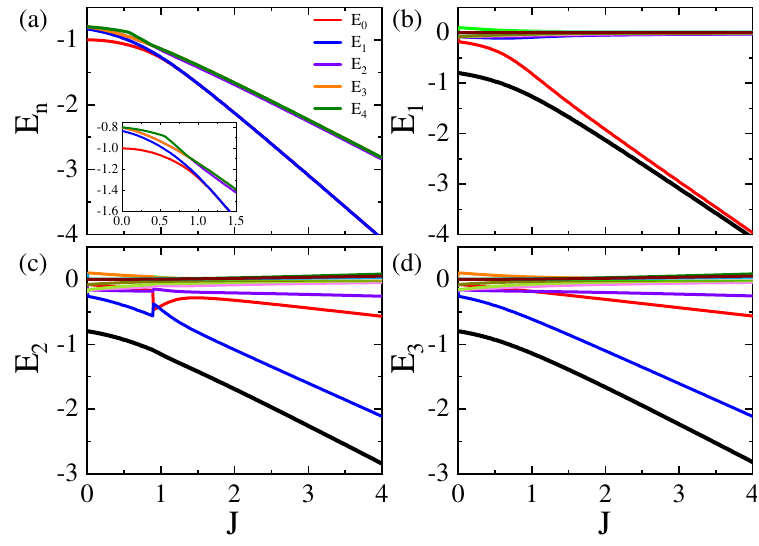}
\caption{(a) The energy levels of the ground state and the first four low-lying excited states for $L=10$. (b), (c) and (d) The pattern components of the first-excited state energy, the second-excited state energy and the third-excited state energy, respectively. The legend of (b), (c) and (d) is consistent with Fig. \ref{fig6} (c1).}\label{fig8}
\end{center}
\end{figure}

We have explored the process of analogue of phase transition in the ground state and analyzed the variation in the contribution of different patterns under different parameters. Now we extend our study to the physics of excited states using the pattern picture, with $L=10$ as an illustrative example. Fig. \ref{fig8} (a) displays the energy spectrum of the ground state and the first four low-lying excited states. Near the Ising interaction strength at $J =1.0$, the ground state and the first excited state become degenerate. As a result, the behavior of pattern components in the first excited state closely resembles that of the ground state, with the pattern $\lambda_1$ dominating in the strong Ising interaction regime, as shown in Fig. \ref{fig8} (b). The significant energy gap between the second/third excited state and the ground state/first excited state suggests that the physics of the highly excited states differs from the ground state. As shown in Fig. \ref{fig8} (c) and (d), the dominant contribution in these higher excited states does not come from the pattern $\lambda_1$, but rather from the pattern $\lambda_{2,3}$. Additionally, the pattern components in the third excited state exhibits a jump around $J=1.0$, as shown in Fig. \ref{fig8} (c), indicating that there occurs something like a first-order phase transition or energy level crossing.

It is well-known that studying excited state physics of a quantum many-body system is more challenging than the ground state, as there are fewer methods for calculating excited states. The pattern picture we propose offers a novel physical picture to study excited state, which will be explored in details in the future.

\section{\label{sec:level5} The Pattern Picture in Larger Lattice Sizes}
In the previous section, we examined pattern behaviors in small lattice sizes, specifically $L = 6, 8, 10$, and $12$. In this section, we extend our analysis to larger lattice sizes, including $L = 16, 32, 64$, and $128$. By studying these various sizes, we can gain valuable insights into the pattern behaviors as the system approaches the thermodynamic limit. 

\begin{figure}[tbp]
\begin{center}
\includegraphics[width = \columnwidth]{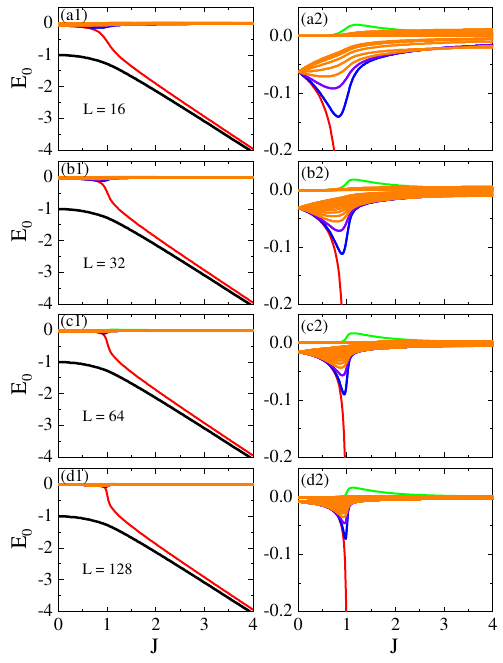}
\caption{Left column: the ground state energy (black solid lines) and the corresponding pattern components (colored solid lines) for (a1) $L = 16$, (b1) $L = 32$, (c1) $L=64$, and (d1) $L=128$. Right column: the corresponding enlarged plots of left column. For clarity, we only mark the pattern $\lambda_1$, the pattern $\lambda_{2,3}$, the pattern $\lambda_{4,5}$ and the pattern $\lambda_{L+1}$ by red, blue, purple and green lines, respectively. All other patterns are marked in orange.}\label{fig9}
\end{center}
\end{figure}

\begin{figure}
\begin{center}
\includegraphics[width = \columnwidth]{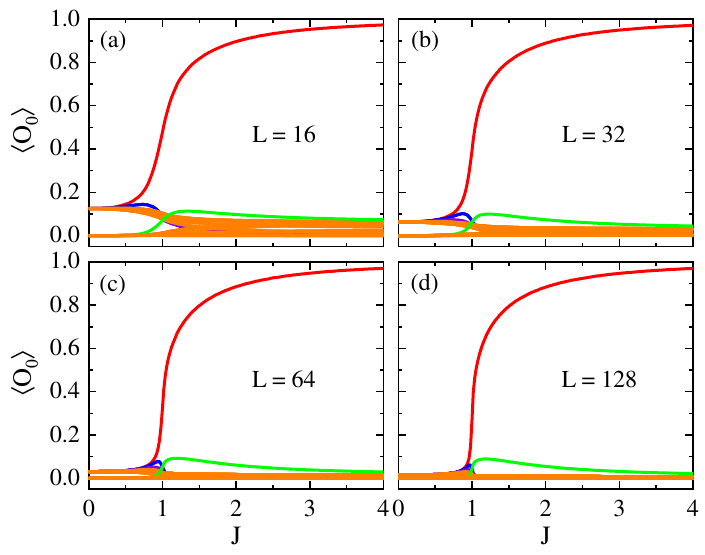}
\caption{The patterns' occupancy of the ground state for lattice sizes (a) $L=16$, (b) $L=32$, (c) $L=64$, and (d) $L=128$. Red, blue, purple and green lines denote the pattern $\lambda_1$, the pattern $\lambda_{2,3}$, the pattern $\lambda_{4,5}$ and the pattern $\lambda_{L+1}$, respectively. Other patterns are marked in orange. }\label{fig10}
\end{center}
\end{figure}

\subsection{\label{sec:level5A} Pattern behaviors of the ground state for lattice sizes $L = 16,32,64,128$}
Figure \ref{fig9} presents the ground state energy (black lines) and the corresponding pattern components (colored lines) for lattice sizes $L = 16, 32, 64$, and $128$. The wavefunctions for these large systems are computed using the density matrix renormalization group (DMRG), and the pattern components are calculated using Eq. (\ref{Ei}). For clarity, we only highlight the pattern $\lambda_1$, the pattern $\lambda_{2,3}$, the pattern $\lambda_{4,5}$ and the pattern $\lambda_{L+1}$ by red, blue, purple and green lines, respectively, while all other patterns are marked in orange. 

A clear comparison between Fig. \ref{fig6} and Fig. \ref{fig9} demonstrates that the microscopic process of the analogue of QPT in the TIM, as discussed for small lattice sizes, remains valid for larger systems, i.e., all patterns in the $\lambda < 0$ group exist in the disordered phase while only the pattern $\lambda_1$ dominates in the ordered phase. However, as the lattice size increases, the pattern changes become sharper, and the turning points progressively toward the point at $J_c = 1$, the QPT point in the thermodynamic limit. Particularly, for the system with $L=128$, the deep dip in the pattern $\lambda_1$ and the sharp peak in the pattern $\lambda_{L+1}$ indicate that the system is approaching the thermodynamic limit. 

Two patterns are particularly noteworthy. The first is the pattern $\lambda_1$, whose energy decreases more rapidly near the phase transition point as the system size increases, clearly reflecting its behaviors in the thermodynamic limit, as shown in the inset of Fig. \ref{fig11} (a1). The second is the pattern $\lambda_{L+1}$, which exhibits a sudden increase near the phase transition point, signifying the role of quantum fluctuations during the phase transition. This observation is significant, as it provides an intuitive understanding of the role played by quantum fluctuations, a perspective that has not been clearly demonstrated in the literature. Similar behaviors are also observed in the patterns' occupancy, calculated using Eq. (\ref{Occ}), as illustrated in Fig. \ref{fig10}. These results exhibit the robustness of the pattern picture in capturing the essential features of the QPT across a wide range of system sizes, from small to the thermodynamic limit.

\subsection{\label{sec:level5B} Determining the QPT point in the TIM using patterns}
It is well-known that the TIM undergoes a second-order QPT from the paramagnetic phase to the ferromagnetic phase, characterized by the divergence of the second-order derivative of the ground state energy at the critical point $J_c=1$ in the thermodynamic limit. The question then arises: how can this phase transition point be determined using the pattern approach?
\begin{figure}
\begin{center}
\includegraphics[width = \columnwidth]{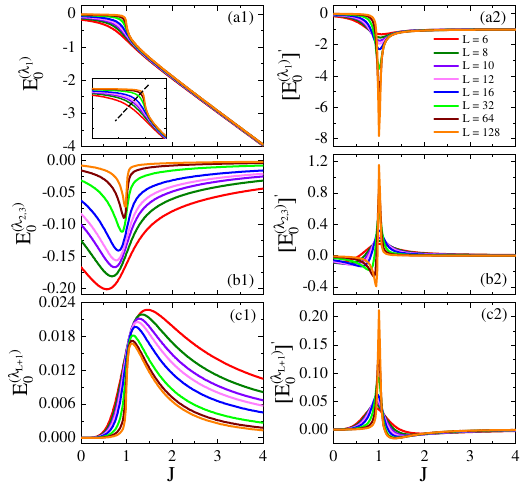}
\caption{The pattern components of the ground state energy for lattice sizes ranging from $L=6$ to $L=128$. Left column: (a1) the pattern $\lambda_1$, (b1) the pattern $\lambda_{2,3}$ and (c1) the pattern $\lambda_{L+1}$. Right column: the first-order derivatives of the pattern components in left column. The QPT occurs at $J_c=1.0$.}\label{fig11}
\end{center}
\end{figure}

Fig. \ref{fig11} shows the pattern components of the ground state energy for lattice sizes ranging from $L=6$ to $L=128$. One can clearly see that the patterns become increasingly sharp near the Ising interaction strength $J=1$ as the system size increases. For simplicity, here we focus on three representative patterns, namely, the pattern $\lambda_1$, the pattern $\lambda_{2,3}$ and the pattern $\lambda_{L+1}$. Other patterns exhibit similar behavior. Surprisingly, when we compute the first-order derivatives of these patterns, we observe a non-analytical point precisely at $J=1$, exactly consistent with the known phase transition point of the TIM, as shown in Fig. \ref{fig11} (a2), (b2), and (c2). 

This phenomenon is particularly intriguing because, while the system as a whole undergoes a continuous (second-order) phase transition, the first-order derivatives of the patterns exhibit a sharp, non-analytical change at the critical point. This suggests that, although the global properties of the system change smoothly during the phase transition, the internal structure of the system undergoes dramatic and abrupt changes. These internal changes are effectively captured by the patterns, providing a novel and intuitive way to identify the critical point. 

\section{\label{Conclusion}Conclusion and Discussion}
We introduce a pattern picture, obtained through two successive diagonalizations, to analyze the well-studied QPT in the TIM. The patterns are constructed from single-body lattice operators weighted by pattern eigenvectors obtained from the first diagonalization. Through the analysis of the energy contributions of the patterns to the ground state as obtained by the second diagonalization, it is shown that different patterns play different roles in the microscopic process of the analogue of QPT from the paramagnetic phase in the weak interaction regime to the ferromagnetic phase in the strong interaction regime. Furthermore, the patterns' occupancy with different Ising interactions uncovers the microscopic process of the analogue of phase transition in an intuitive way. Remarkably, the critical point is precisely identified by the first-order derivatives of the pattern components of the ground state energy.

Although the one-dimensional TIM can be exactly solved using the Jordan-Winger transformation, the pattern picture is not confined to exactly solvable model. We have successfully extended this approach to other spin models, including the one-dimensional TIM with a longitudinal field \cite{Yang2023a}, the one-dimensional axial next-nearest-neighbor Ising model in a transverse field \cite{Yang2023c} and the one-dimensional anisotropic quantum XY model with a transverse field \cite{Yang2023d}. For two-dimensional systems, we have applied the pattern picture to frustrated models such as the $J_1$-$J_2$ Heisenberg model on the square lattice\cite{Yang2023e} and the Shastry-Sutherland model \cite{Yang2024a}. Extensions to other models exhibiting phase transitions and comparisons to conventional formulations of phase transitions are left for future works.

Additionally, due to its simplicity and fundamental nature, the TIM holds significant practical relevance for quantum simulations \cite{Georgescu2014}. These experimental platforms may provide an opportunity to test our pattern picture, particularly in characterizing the energy competitions between different patterns and elucidating their roles in the microscopic process of QPT. 

\section{Acknowledgments}
The work is partly supported by the National Key Research and Development Program of China (Grant No. 2022YFA1402704) and the programs for NSFC of China (Grant No. 12247101).


\begin{thebibliography}{34}%
\makeatletter
\providecommand \@ifxundefined [1]{%
 \@ifx{#1\undefined}
}%
\providecommand \@ifnum [1]{%
 \ifnum #1\expandafter \@firstoftwo
 \else \expandafter \@secondoftwo
 \fi
}%
\providecommand \@ifx [1]{%
 \ifx #1\expandafter \@firstoftwo
 \else \expandafter \@secondoftwo
 \fi
}%
\providecommand \natexlab [1]{#1}%
\providecommand \enquote  [1]{``#1''}%
\providecommand \bibnamefont  [1]{#1}%
\providecommand \bibfnamefont [1]{#1}%
\providecommand \citenamefont [1]{#1}%
\providecommand \href@noop [0]{\@secondoftwo}%
\providecommand \href [0]{\begingroup \@sanitize@url \@href}%
\providecommand \@href[1]{\@@startlink{#1}\@@href}%
\providecommand \@@href[1]{\endgroup#1\@@endlink}%
\providecommand \@sanitize@url [0]{\catcode `\\12\catcode `\$12\catcode
  `\&12\catcode `\#12\catcode `\^12\catcode `\_12\catcode `\%12\relax}%
\providecommand \@@startlink[1]{}%
\providecommand \@@endlink[0]{}%
\providecommand \url  [0]{\begingroup\@sanitize@url \@url }%
\providecommand \@url [1]{\endgroup\@href {#1}{\urlprefix }}%
\providecommand \urlprefix  [0]{URL }%
\providecommand \Eprint [0]{\href }%
\providecommand \doibase [0]{https://doi.org/}%
\providecommand \selectlanguage [0]{\@gobble}%
\providecommand \bibinfo  [0]{\@secondoftwo}%
\providecommand \bibfield  [0]{\@secondoftwo}%
\providecommand \translation [1]{[#1]}%
\providecommand \BibitemOpen [0]{}%
\providecommand \bibitemStop [0]{}%
\providecommand \bibitemNoStop [0]{.\EOS\space}%
\providecommand \EOS [0]{\spacefactor3000\relax}%
\providecommand \BibitemShut  [1]{\csname bibitem#1\endcsname}%
\let\auto@bib@innerbib\@empty
\bibitem [{\citenamefont {Landau}(1937)}]{Landau1937}%
  \BibitemOpen
  \bibfield  {author} {\bibinfo {author} {\bibfnamefont {L.~D.}\ \bibnamefont
  {Landau}},\ }\bibfield  {title} {\bibinfo {title} {On the theory of phase
  transitions. i.},\ }\href@noop {} {\bibfield  {journal} {\bibinfo  {journal}
  {Phys. Z. Sowjet.}\ }\textbf {\bibinfo {volume} {11}} (\bibinfo {year}
  {1937})}\BibitemShut {NoStop}%
\bibitem [{\citenamefont {Landau}\ and\ \citenamefont
  {Lifshitz}(1980)}]{Landau1980}%
  \BibitemOpen
  \bibfield  {author} {\bibinfo {author} {\bibfnamefont {L.~D.}\ \bibnamefont
  {Landau}}\ and\ \bibinfo {author} {\bibfnamefont {E.~M.}\ \bibnamefont
  {Lifshitz}},\ }\href@noop {} {\emph {\bibinfo {title} {Course of Theoretical
  Physics Vol. 5 Statistical Physics Part I $\&$ II 3rd ed.}}}\ (\bibinfo
  {publisher} {Elsevier Ltd.},\ \bibinfo {address} {Amsterdam},\ \bibinfo
  {year} {1980})\BibitemShut {NoStop}%
\bibitem [{\citenamefont {Wilson}\ and\ \citenamefont
  {Kogut}(1974)}]{Wilson1974}%
  \BibitemOpen
  \bibfield  {author} {\bibinfo {author} {\bibfnamefont {K.~G.}\ \bibnamefont
  {Wilson}}\ and\ \bibinfo {author} {\bibfnamefont {J.}~\bibnamefont {Kogut}},\
  }\bibfield  {title} {\bibinfo {title} {The renormalization group and the
  $\epsilon$ expansion},\ }\href@noop {} {\bibfield  {journal} {\bibinfo
  {journal} {Phys. Rep.}\ }\textbf {\bibinfo {volume} {12}},\ \bibinfo {pages}
  {75} (\bibinfo {year} {1974})}\BibitemShut {NoStop}%
\bibitem [{\citenamefont {Chaikin}\ and\ \citenamefont
  {Lubensky}(2000)}]{Chaikin2000}%
  \BibitemOpen
  \bibfield  {author} {\bibinfo {author} {\bibfnamefont {P.~M.}\ \bibnamefont
  {Chaikin}}\ and\ \bibinfo {author} {\bibfnamefont {T.~C.}\ \bibnamefont
  {Lubensky}},\ }\href@noop {} {\emph {\bibinfo {title} {Principles of
  Condensed Matter Physics}}}\ (\bibinfo  {publisher} {Cambridge University
  Press},\ \bibinfo {year} {2000})\BibitemShut {NoStop}%
\bibitem [{\citenamefont {Kastner}(2008)}]{Kastner2008}%
  \BibitemOpen
  \bibfield  {author} {\bibinfo {author} {\bibfnamefont {M.}~\bibnamefont
  {Kastner}},\ }\bibfield  {title} {\bibinfo {title} {Phase transitions and
  configuration space topology},\ }\href
  {https://doi.org/10.1103/RevModPhys.80.167} {\bibfield  {journal} {\bibinfo
  {journal} {Rev. Mod. Phys.}\ }\textbf {\bibinfo {volume} {80}},\ \bibinfo
  {pages} {167} (\bibinfo {year} {2008})}\BibitemShut {NoStop}%
\bibitem [{\citenamefont {Yang}\ and\ \citenamefont {Lee}(1952)}]{Yang1952}%
  \BibitemOpen
  \bibfield  {author} {\bibinfo {author} {\bibfnamefont {C.~N.}\ \bibnamefont
  {Yang}}\ and\ \bibinfo {author} {\bibfnamefont {T.~D.}\ \bibnamefont {Lee}},\
  }\bibfield  {title} {\bibinfo {title} {Statistical theory of equations of
  state and phase transitions. i. theory of condensation},\ }\href
  {https://doi.org/10.1103/PhysRev.87.404} {\bibfield  {journal} {\bibinfo
  {journal} {Phys. Rev.}\ }\textbf {\bibinfo {volume} {87}},\ \bibinfo {pages}
  {404} (\bibinfo {year} {1952})}\BibitemShut {NoStop}%
\bibitem [{\citenamefont {Kibble}(1976)}]{Kibble1976}%
  \BibitemOpen
  \bibfield  {author} {\bibinfo {author} {\bibfnamefont {T.~W.~B.}\
  \bibnamefont {Kibble}},\ }\bibfield  {title} {\bibinfo {title} {Topology of
  cosmic domains and strings},\ }\href
  {https://doi.org/10.1088/0305-4470/9/8/029} {\bibfield  {journal} {\bibinfo
  {journal} {Journal of Physics A: Mathematical and General}\ }\textbf
  {\bibinfo {volume} {9}},\ \bibinfo {pages} {1387} (\bibinfo {year}
  {1976})}\BibitemShut {NoStop}%
\bibitem [{\citenamefont {Zurek}(1985)}]{Zurek1985}%
  \BibitemOpen
  \bibfield  {author} {\bibinfo {author} {\bibfnamefont {W.}~\bibnamefont
  {Zurek}},\ }\bibfield  {title} {\bibinfo {title} {Cosmological experiments in
  superfluid helium?},\ }\href@noop {} {\bibfield  {journal} {\bibinfo
  {journal} {Nature}\ }\textbf {\bibinfo {volume} {317}},\ \bibinfo {pages}
  {505} (\bibinfo {year} {1985})}\BibitemShut {NoStop}%
\bibitem [{\citenamefont {Zurek}(1996)}]{Zurek1996}%
  \BibitemOpen
  \bibfield  {author} {\bibinfo {author} {\bibfnamefont {W.}~\bibnamefont
  {Zurek}},\ }\bibfield  {title} {\bibinfo {title} {Cosmological experiments in
  condensed matter systems},\ }\href
  {https://doi.org/https://doi.org/10.1016/S0370-1573(96)00009-9} {\bibfield
  {journal} {\bibinfo  {journal} {Physics Reports}\ }\textbf {\bibinfo {volume}
  {276}},\ \bibinfo {pages} {177} (\bibinfo {year} {1996})}\BibitemShut
  {NoStop}%
\bibitem [{\citenamefont {Weiler}\ \emph {et~al.}(2008)\citenamefont {Weiler},
  \citenamefont {Neely}, \citenamefont {Scherer}, \citenamefont {Bradley},
  \citenamefont {Davis},\ and\ \citenamefont {Anderson}}]{Weiler2008}%
  \BibitemOpen
  \bibfield  {author} {\bibinfo {author} {\bibfnamefont {C.~N.}\ \bibnamefont
  {Weiler}}, \bibinfo {author} {\bibfnamefont {T.~W.}\ \bibnamefont {Neely}},
  \bibinfo {author} {\bibfnamefont {D.~R.}\ \bibnamefont {Scherer}}, \bibinfo
  {author} {\bibfnamefont {A.~S.}\ \bibnamefont {Bradley}}, \bibinfo {author}
  {\bibfnamefont {M.~J.}\ \bibnamefont {Davis}},\ and\ \bibinfo {author}
  {\bibfnamefont {B.~P.}\ \bibnamefont {Anderson}},\ }\bibfield  {title}
  {\bibinfo {title} {Spontaneous vortices in the formation of bose–einstein
  condensates},\ }\href {https://doi.org/10.1038/nature07334} {\bibfield
  {journal} {\bibinfo  {journal} {Nature}\ }\textbf {\bibinfo {volume} {455}},\
  \bibinfo {pages} {948} (\bibinfo {year} {2008})}\BibitemShut {NoStop}%
\bibitem [{\citenamefont {Pyka}\ \emph {et~al.}(2013)\citenamefont {Pyka},
  \citenamefont {Keller}, \citenamefont {Partner}, \citenamefont {Nigmatullin},
  \citenamefont {Burgermeister}, \citenamefont {Meier}, \citenamefont
  {Kuhlmann}, \citenamefont {Retzker}, \citenamefont {Plenio}, \citenamefont
  {Zurek}, \citenamefont {del Campo},\ and\ \citenamefont
  {Mehlstäubler}}]{Pyka2013}%
  \BibitemOpen
  \bibfield  {author} {\bibinfo {author} {\bibfnamefont {K.}~\bibnamefont
  {Pyka}}, \bibinfo {author} {\bibfnamefont {J.}~\bibnamefont {Keller}},
  \bibinfo {author} {\bibfnamefont {H.~L.}\ \bibnamefont {Partner}}, \bibinfo
  {author} {\bibfnamefont {R.}~\bibnamefont {Nigmatullin}}, \bibinfo {author}
  {\bibfnamefont {T.}~\bibnamefont {Burgermeister}}, \bibinfo {author}
  {\bibfnamefont {D.~M.}\ \bibnamefont {Meier}}, \bibinfo {author}
  {\bibfnamefont {K.}~\bibnamefont {Kuhlmann}}, \bibinfo {author}
  {\bibfnamefont {A.}~\bibnamefont {Retzker}}, \bibinfo {author} {\bibfnamefont
  {M.~B.}\ \bibnamefont {Plenio}}, \bibinfo {author} {\bibfnamefont {W.~H.}\
  \bibnamefont {Zurek}}, \bibinfo {author} {\bibfnamefont {A.}~\bibnamefont
  {del Campo}},\ and\ \bibinfo {author} {\bibfnamefont {T.~E.}\ \bibnamefont
  {Mehlstäubler}},\ }\bibfield  {title} {\bibinfo {title} {Topological defect
  formation and spontaneous symmetry breaking in ion coulomb crystals},\ }\href
  {https://doi.org/10.1038/ncomms3291} {\bibfield  {journal} {\bibinfo
  {journal} {Nature Communications}\ }\textbf {\bibinfo {volume} {4}},\
  \bibinfo {pages} {2291} (\bibinfo {year} {2013})}\BibitemShut {NoStop}%
\bibitem [{\citenamefont {Deutschländer}\ \emph {et~al.}(2015)\citenamefont
  {Deutschländer}, \citenamefont {Dillmann}, \citenamefont {Maret},\ and\
  \citenamefont {Keim}}]{Deutschlander2015}%
  \BibitemOpen
  \bibfield  {author} {\bibinfo {author} {\bibfnamefont {S.}~\bibnamefont
  {Deutschländer}}, \bibinfo {author} {\bibfnamefont {P.}~\bibnamefont
  {Dillmann}}, \bibinfo {author} {\bibfnamefont {G.}~\bibnamefont {Maret}},\
  and\ \bibinfo {author} {\bibfnamefont {P.}~\bibnamefont {Keim}},\ }\bibfield
  {title} {\bibinfo {title} {Kibble–zurek mechanism in colloidal
  monolayers},\ }\href {https://doi.org/10.1073/pnas.1500763112} {\bibfield
  {journal} {\bibinfo  {journal} {Proceedings of the National Academy of
  Sciences}\ }\textbf {\bibinfo {volume} {112}},\ \bibinfo {pages} {6925}
  (\bibinfo {year} {2015})}\BibitemShut {NoStop}%
\bibitem [{\citenamefont {Keesling}\ \emph {et~al.}(2019)\citenamefont
  {Keesling}, \citenamefont {Omran}, \citenamefont {Levine}, \citenamefont
  {Bernien}, \citenamefont {Pichler}, \citenamefont {Choi}, \citenamefont
  {Samajdar}, \citenamefont {Schwartz}, \citenamefont {Silvi}, \citenamefont
  {Sachdev}, \citenamefont {Zoller}, \citenamefont {Endres}, \citenamefont
  {Greiner}, \citenamefont {Vuletić},\ and\ \citenamefont
  {Lukin}}]{Keesling2019}%
  \BibitemOpen
  \bibfield  {author} {\bibinfo {author} {\bibfnamefont {A.}~\bibnamefont
  {Keesling}}, \bibinfo {author} {\bibfnamefont {A.}~\bibnamefont {Omran}},
  \bibinfo {author} {\bibfnamefont {H.}~\bibnamefont {Levine}}, \bibinfo
  {author} {\bibfnamefont {H.}~\bibnamefont {Bernien}}, \bibinfo {author}
  {\bibfnamefont {H.}~\bibnamefont {Pichler}}, \bibinfo {author} {\bibfnamefont
  {S.}~\bibnamefont {Choi}}, \bibinfo {author} {\bibfnamefont {R.}~\bibnamefont
  {Samajdar}}, \bibinfo {author} {\bibfnamefont {S.}~\bibnamefont {Schwartz}},
  \bibinfo {author} {\bibfnamefont {P.}~\bibnamefont {Silvi}}, \bibinfo
  {author} {\bibfnamefont {S.}~\bibnamefont {Sachdev}}, \bibinfo {author}
  {\bibfnamefont {P.}~\bibnamefont {Zoller}}, \bibinfo {author} {\bibfnamefont
  {M.}~\bibnamefont {Endres}}, \bibinfo {author} {\bibfnamefont
  {M.}~\bibnamefont {Greiner}}, \bibinfo {author} {\bibfnamefont
  {V.}~\bibnamefont {Vuletić}},\ and\ \bibinfo {author} {\bibfnamefont
  {M.~D.}\ \bibnamefont {Lukin}},\ }\bibfield  {title} {\bibinfo {title}
  {Quantum kibble–zurek mechanism and critical dynamics on a programmable
  rydberg simulator},\ }\href {https://doi.org/10.1038/s41586-019-1070-1}
  {\bibfield  {journal} {\bibinfo  {journal} {Nature}\ }\textbf {\bibinfo
  {volume} {568}},\ \bibinfo {pages} {207} (\bibinfo {year}
  {2019})}\BibitemShut {NoStop}%
\bibitem [{\citenamefont {Ko}\ \emph {et~al.}(2019)\citenamefont {Ko},
  \citenamefont {Park},\ and\ \citenamefont {Shin}}]{Ko2019}%
  \BibitemOpen
  \bibfield  {author} {\bibinfo {author} {\bibfnamefont {B.}~\bibnamefont
  {Ko}}, \bibinfo {author} {\bibfnamefont {J.~W.}\ \bibnamefont {Park}},\ and\
  \bibinfo {author} {\bibfnamefont {Y.}~\bibnamefont {Shin}},\ }\bibfield
  {title} {\bibinfo {title} {Kibble–zurek universality in a strongly
  interacting fermi superfluid},\ }\href
  {https://doi.org/10.1038/s41567-019-0650-1} {\bibfield  {journal} {\bibinfo
  {journal} {Nature Physics}\ }\textbf {\bibinfo {volume} {15}},\ \bibinfo
  {pages} {1227} (\bibinfo {year} {2019})}\BibitemShut {NoStop}%
\bibitem [{\citenamefont {Schmitt}\ \emph {et~al.}(2022)\citenamefont
  {Schmitt}, \citenamefont {Rams}, \citenamefont {Dziarmaga}, \citenamefont
  {Heyl},\ and\ \citenamefont {Zurek}}]{Schmitt2022}%
  \BibitemOpen
  \bibfield  {author} {\bibinfo {author} {\bibfnamefont {M.}~\bibnamefont
  {Schmitt}}, \bibinfo {author} {\bibfnamefont {M.~M.}\ \bibnamefont {Rams}},
  \bibinfo {author} {\bibfnamefont {J.}~\bibnamefont {Dziarmaga}}, \bibinfo
  {author} {\bibfnamefont {M.}~\bibnamefont {Heyl}},\ and\ \bibinfo {author}
  {\bibfnamefont {W.~H.}\ \bibnamefont {Zurek}},\ }\bibfield  {title} {\bibinfo
  {title} {Quantum phase transition dynamics in the two-dimensional
  transverse-field ising model},\ }\href
  {https://doi.org/10.1126/sciadv.abl6850} {\bibfield  {journal} {\bibinfo
  {journal} {Science Advances}\ }\textbf {\bibinfo {volume} {8}},\ \bibinfo
  {pages} {eabl6850} (\bibinfo {year} {2022})}\BibitemShut {NoStop}%
\bibitem [{\citenamefont {Braun}\ \emph {et~al.}(2015)\citenamefont {Braun},
  \citenamefont {Friesdorf}, \citenamefont {Hodgman}, \citenamefont
  {Schreiber}, \citenamefont {Ronzheimer}, \citenamefont {Riera}, \citenamefont
  {del Rey}, \citenamefont {Bloch}, \citenamefont {Eisert},\ and\ \citenamefont
  {Schneider}}]{Braun2015}%
  \BibitemOpen
  \bibfield  {author} {\bibinfo {author} {\bibfnamefont {S.}~\bibnamefont
  {Braun}}, \bibinfo {author} {\bibfnamefont {M.}~\bibnamefont {Friesdorf}},
  \bibinfo {author} {\bibfnamefont {S.~S.}\ \bibnamefont {Hodgman}}, \bibinfo
  {author} {\bibfnamefont {M.}~\bibnamefont {Schreiber}}, \bibinfo {author}
  {\bibfnamefont {J.~P.}\ \bibnamefont {Ronzheimer}}, \bibinfo {author}
  {\bibfnamefont {A.}~\bibnamefont {Riera}}, \bibinfo {author} {\bibfnamefont
  {M.}~\bibnamefont {del Rey}}, \bibinfo {author} {\bibfnamefont
  {I.}~\bibnamefont {Bloch}}, \bibinfo {author} {\bibfnamefont
  {J.}~\bibnamefont {Eisert}},\ and\ \bibinfo {author} {\bibfnamefont
  {U.}~\bibnamefont {Schneider}},\ }\bibfield  {title} {\bibinfo {title}
  {Emergence of coherence and the dynamics of quantum phase transitions},\
  }\href {https://doi.org/10.1073/pnas.1408861112} {\bibfield  {journal}
  {\bibinfo  {journal} {Proceedings of the National Academy of Sciences}\
  }\textbf {\bibinfo {volume} {112}},\ \bibinfo {pages} {3641} (\bibinfo {year}
  {2015})}\BibitemShut {NoStop}%
\bibitem [{\citenamefont {Yang}\ and\ \citenamefont {Luo}(2024)}]{Yang2022b}%
  \BibitemOpen
  \bibfield  {author} {\bibinfo {author} {\bibfnamefont {Y.-T.}\ \bibnamefont
  {Yang}}\ and\ \bibinfo {author} {\bibfnamefont {H.-G.}\ \bibnamefont {Luo}},\
  }\bibfield  {title} {\bibinfo {title} {Dissecting superradiant phase
  transition in the quantum rabi model},\ }\href
  {https://doi.org/10.1088/0256-307X/41/12/120501} {\bibfield  {journal}
  {\bibinfo  {journal} {Chin. Phys. Lett.}\ }\textbf {\bibinfo {volume} {41}},\
  \bibinfo {pages} {120501} (\bibinfo {year} {2024})}\BibitemShut {NoStop}%
\bibitem [{\citenamefont {Sachdev}(2011)}]{Sachdev2011}%
  \BibitemOpen
  \bibfield  {author} {\bibinfo {author} {\bibfnamefont {S.}~\bibnamefont
  {Sachdev}},\ }\href@noop {} {\emph {\bibinfo {title} {Quantum phase
  transitions}}},\ \bibinfo {edition} {second ed.}\ ed.\ (\bibinfo  {publisher}
  {Cambridge University Press},\ \bibinfo {address} {Cambridge},\ \bibinfo
  {year} {2011})\BibitemShut {NoStop}%
\bibitem [{\citenamefont {Dutta}\ \emph {et~al.}(2015)\citenamefont {Dutta},
  \citenamefont {Aeppli}, \citenamefont {Chakrabarti}, \citenamefont
  {Divakaran}, \citenamefont {Rosenbaum},\ and\ \citenamefont
  {Sen}}]{Dutta2015}%
  \BibitemOpen
  \bibfield  {author} {\bibinfo {author} {\bibfnamefont {A.}~\bibnamefont
  {Dutta}}, \bibinfo {author} {\bibfnamefont {G.}~\bibnamefont {Aeppli}},
  \bibinfo {author} {\bibfnamefont {B.~K.}\ \bibnamefont {Chakrabarti}},
  \bibinfo {author} {\bibfnamefont {U.}~\bibnamefont {Divakaran}}, \bibinfo
  {author} {\bibfnamefont {T.~F.}\ \bibnamefont {Rosenbaum}},\ and\ \bibinfo
  {author} {\bibfnamefont {D.}~\bibnamefont {Sen}},\ }in\ \href@noop {} {\emph
  {\bibinfo {booktitle} {Quantum Phase Transitions in Transverse Field Spin
  Models: From Statistical Physics to Quantum Information}}}\ (\bibinfo
  {publisher} {Cambridge University Press},\ \bibinfo {year}
  {2015})\BibitemShut {NoStop}%
\bibitem [{\citenamefont {Coldea}\ \emph {et~al.}(2010)\citenamefont {Coldea},
  \citenamefont {Tennant}, \citenamefont {Wheeler}, \citenamefont {Wawrzynska},
  \citenamefont {Prabhakaran}, \citenamefont {Telling}, \citenamefont
  {Habicht}, \citenamefont {Smeibidl},\ and\ \citenamefont
  {Kiefer}}]{Coldea2010}%
  \BibitemOpen
  \bibfield  {author} {\bibinfo {author} {\bibfnamefont {R.}~\bibnamefont
  {Coldea}}, \bibinfo {author} {\bibfnamefont {D.~A.}\ \bibnamefont {Tennant}},
  \bibinfo {author} {\bibfnamefont {E.~M.}\ \bibnamefont {Wheeler}}, \bibinfo
  {author} {\bibfnamefont {E.}~\bibnamefont {Wawrzynska}}, \bibinfo {author}
  {\bibfnamefont {D.}~\bibnamefont {Prabhakaran}}, \bibinfo {author}
  {\bibfnamefont {M.}~\bibnamefont {Telling}}, \bibinfo {author} {\bibfnamefont
  {K.}~\bibnamefont {Habicht}}, \bibinfo {author} {\bibfnamefont
  {P.}~\bibnamefont {Smeibidl}},\ and\ \bibinfo {author} {\bibfnamefont
  {K.}~\bibnamefont {Kiefer}},\ }\bibfield  {title} {\bibinfo {title} {Quantum
  criticality in an ising chain: Experimental evidence for emergent e$_8$
  symmetry},\ }\href {https://doi.org/10.1126/science.1180085} {\bibfield
  {journal} {\bibinfo  {journal} {Science}\ }\textbf {\bibinfo {volume}
  {327}},\ \bibinfo {pages} {177} (\bibinfo {year} {2010})}\BibitemShut
  {NoStop}%
\bibitem [{\citenamefont {Breunig}\ \emph {et~al.}(2017)\citenamefont
  {Breunig}, \citenamefont {Garst}, \citenamefont {Klümper}, \citenamefont
  {Rohrkamp}, \citenamefont {Turnbull},\ and\ \citenamefont
  {Lorenz}}]{Breunig2017}%
  \BibitemOpen
  \bibfield  {author} {\bibinfo {author} {\bibfnamefont {O.}~\bibnamefont
  {Breunig}}, \bibinfo {author} {\bibfnamefont {M.}~\bibnamefont {Garst}},
  \bibinfo {author} {\bibfnamefont {A.}~\bibnamefont {Klümper}}, \bibinfo
  {author} {\bibfnamefont {J.}~\bibnamefont {Rohrkamp}}, \bibinfo {author}
  {\bibfnamefont {M.~M.}\ \bibnamefont {Turnbull}},\ and\ \bibinfo {author}
  {\bibfnamefont {T.}~\bibnamefont {Lorenz}},\ }\bibfield  {title} {\bibinfo
  {title} {Quantum criticality in the spin-1/2 heisenberg chain system copper
  pyrazine dinitrate},\ }\href {https://doi.org/10.1126/sciadv.aao3773}
  {\bibfield  {journal} {\bibinfo  {journal} {Science Advances}\ }\textbf
  {\bibinfo {volume} {3}},\ \bibinfo {pages} {eaao3773} (\bibinfo {year}
  {2017})}\BibitemShut {NoStop}%
\bibitem [{\citenamefont {Friedenauer}\ \emph {et~al.}(2008)\citenamefont
  {Friedenauer}, \citenamefont {Schmitz}, \citenamefont {Glueckert},
  \citenamefont {Porras},\ and\ \citenamefont {Schaetz}}]{Friedenauer2008}%
  \BibitemOpen
  \bibfield  {author} {\bibinfo {author} {\bibfnamefont {A.}~\bibnamefont
  {Friedenauer}}, \bibinfo {author} {\bibfnamefont {H.}~\bibnamefont
  {Schmitz}}, \bibinfo {author} {\bibfnamefont {J.~T.}\ \bibnamefont
  {Glueckert}}, \bibinfo {author} {\bibfnamefont {D.}~\bibnamefont {Porras}},\
  and\ \bibinfo {author} {\bibfnamefont {T.}~\bibnamefont {Schaetz}},\
  }\bibfield  {title} {\bibinfo {title} {Simulating a quantum magnet with
  trapped ions},\ }\href {https://doi.org/10.1038/nphys1032} {\bibfield
  {journal} {\bibinfo  {journal} {Nature Physics}\ }\textbf {\bibinfo {volume}
  {4}},\ \bibinfo {pages} {757} (\bibinfo {year} {2008})}\BibitemShut {NoStop}%
\bibitem [{\citenamefont {Islam}\ \emph {et~al.}(2011)\citenamefont {Islam},
  \citenamefont {Edwards}, \citenamefont {Kim}, \citenamefont {Korenblit},
  \citenamefont {Noh}, \citenamefont {Carmichael}, \citenamefont {Lin},
  \citenamefont {Duan}, \citenamefont {Joseph~Wang}, \citenamefont
  {Freericks},\ and\ \citenamefont {Monroe}}]{Islam2011}%
  \BibitemOpen
  \bibfield  {author} {\bibinfo {author} {\bibfnamefont {R.}~\bibnamefont
  {Islam}}, \bibinfo {author} {\bibfnamefont {E.}~\bibnamefont {Edwards}},
  \bibinfo {author} {\bibfnamefont {K.}~\bibnamefont {Kim}}, \bibinfo {author}
  {\bibfnamefont {S.}~\bibnamefont {Korenblit}}, \bibinfo {author}
  {\bibfnamefont {C.}~\bibnamefont {Noh}}, \bibinfo {author} {\bibfnamefont
  {H.}~\bibnamefont {Carmichael}}, \bibinfo {author} {\bibfnamefont {G.-D.}\
  \bibnamefont {Lin}}, \bibinfo {author} {\bibfnamefont {L.-M.}\ \bibnamefont
  {Duan}}, \bibinfo {author} {\bibfnamefont {C.-C.}\ \bibnamefont
  {Joseph~Wang}}, \bibinfo {author} {\bibfnamefont {J.}~\bibnamefont
  {Freericks}},\ and\ \bibinfo {author} {\bibfnamefont {C.}~\bibnamefont
  {Monroe}},\ }\bibfield  {title} {\bibinfo {title} {Onset of a quantum phase
  transition with a trapped ion quantum simulator},\ }\href
  {https://doi.org/10.1038/ncomms1374} {\bibfield  {journal} {\bibinfo
  {journal} {Nature Communications}\ }\textbf {\bibinfo {volume} {2}},\
  \bibinfo {pages} {377} (\bibinfo {year} {2011})}\BibitemShut {NoStop}%
\bibitem [{\citenamefont {Kim}\ \emph {et~al.}(2011)\citenamefont {Kim},
  \citenamefont {Korenblit}, \citenamefont {Islam}, \citenamefont {Edwards},
  \citenamefont {Chang}, \citenamefont {Noh}, \citenamefont {Carmichael},
  \citenamefont {Lin}, \citenamefont {Duan}, \citenamefont {Wang},
  \citenamefont {Freericks},\ and\ \citenamefont {Monroe}}]{Kim2011}%
  \BibitemOpen
  \bibfield  {author} {\bibinfo {author} {\bibfnamefont {K.}~\bibnamefont
  {Kim}}, \bibinfo {author} {\bibfnamefont {S.}~\bibnamefont {Korenblit}},
  \bibinfo {author} {\bibfnamefont {R.}~\bibnamefont {Islam}}, \bibinfo
  {author} {\bibfnamefont {E.~E.}\ \bibnamefont {Edwards}}, \bibinfo {author}
  {\bibfnamefont {M.-S.}\ \bibnamefont {Chang}}, \bibinfo {author}
  {\bibfnamefont {C.}~\bibnamefont {Noh}}, \bibinfo {author} {\bibfnamefont
  {H.}~\bibnamefont {Carmichael}}, \bibinfo {author} {\bibfnamefont {G.-D.}\
  \bibnamefont {Lin}}, \bibinfo {author} {\bibfnamefont {L.-M.}\ \bibnamefont
  {Duan}}, \bibinfo {author} {\bibfnamefont {C.~C.~J.}\ \bibnamefont {Wang}},
  \bibinfo {author} {\bibfnamefont {J.~K.}\ \bibnamefont {Freericks}},\ and\
  \bibinfo {author} {\bibfnamefont {C.}~\bibnamefont {Monroe}},\ }\bibfield
  {title} {\bibinfo {title} {Quantum simulation of the transverse ising model
  with trapped ions},\ }\href {https://doi.org/10.1088/1367-2630/13/10/105003}
  {\bibfield  {journal} {\bibinfo  {journal} {New Journal of Physics}\ }\textbf
  {\bibinfo {volume} {13}},\ \bibinfo {pages} {105003} (\bibinfo {year}
  {2011})}\BibitemShut {NoStop}%
\bibitem [{\citenamefont {Johnson}\ \emph {et~al.}(2011)\citenamefont
  {Johnson}, \citenamefont {Amin}, \citenamefont {Gildert}, \citenamefont
  {Lanting}, \citenamefont {Hamze}, \citenamefont {Dickson}, \citenamefont
  {Harris}, \citenamefont {Berkley}, \citenamefont {Johansson}, \citenamefont
  {Bunyk}, \citenamefont {Chapple}, \citenamefont {Enderud}, \citenamefont
  {Hilton}, \citenamefont {Karimi}, \citenamefont {Ladizinsky}, \citenamefont
  {Ladizinsky}, \citenamefont {Oh}, \citenamefont {Perminov}, \citenamefont
  {Rich}, \citenamefont {Thom}, \citenamefont {Tolkacheva}, \citenamefont
  {Truncik}, \citenamefont {Uchaikin}, \citenamefont {Wang}, \citenamefont
  {Wilson},\ and\ \citenamefont {Rose}}]{Johnson2011}%
  \BibitemOpen
  \bibfield  {author} {\bibinfo {author} {\bibfnamefont {M.~W.}\ \bibnamefont
  {Johnson}}, \bibinfo {author} {\bibfnamefont {M.~H.~S.}\ \bibnamefont
  {Amin}}, \bibinfo {author} {\bibfnamefont {S.}~\bibnamefont {Gildert}},
  \bibinfo {author} {\bibfnamefont {T.}~\bibnamefont {Lanting}}, \bibinfo
  {author} {\bibfnamefont {F.}~\bibnamefont {Hamze}}, \bibinfo {author}
  {\bibfnamefont {N.}~\bibnamefont {Dickson}}, \bibinfo {author} {\bibfnamefont
  {R.}~\bibnamefont {Harris}}, \bibinfo {author} {\bibfnamefont {A.~J.}\
  \bibnamefont {Berkley}}, \bibinfo {author} {\bibfnamefont {J.}~\bibnamefont
  {Johansson}}, \bibinfo {author} {\bibfnamefont {P.}~\bibnamefont {Bunyk}},
  \bibinfo {author} {\bibfnamefont {E.~M.}\ \bibnamefont {Chapple}}, \bibinfo
  {author} {\bibfnamefont {C.}~\bibnamefont {Enderud}}, \bibinfo {author}
  {\bibfnamefont {J.~P.}\ \bibnamefont {Hilton}}, \bibinfo {author}
  {\bibfnamefont {K.}~\bibnamefont {Karimi}}, \bibinfo {author} {\bibfnamefont
  {E.}~\bibnamefont {Ladizinsky}}, \bibinfo {author} {\bibfnamefont
  {N.}~\bibnamefont {Ladizinsky}}, \bibinfo {author} {\bibfnamefont
  {T.}~\bibnamefont {Oh}}, \bibinfo {author} {\bibfnamefont {I.}~\bibnamefont
  {Perminov}}, \bibinfo {author} {\bibfnamefont {C.}~\bibnamefont {Rich}},
  \bibinfo {author} {\bibfnamefont {M.~C.}\ \bibnamefont {Thom}}, \bibinfo
  {author} {\bibfnamefont {E.}~\bibnamefont {Tolkacheva}}, \bibinfo {author}
  {\bibfnamefont {C.~J.~S.}\ \bibnamefont {Truncik}}, \bibinfo {author}
  {\bibfnamefont {S.}~\bibnamefont {Uchaikin}}, \bibinfo {author}
  {\bibfnamefont {J.}~\bibnamefont {Wang}}, \bibinfo {author} {\bibfnamefont
  {B.}~\bibnamefont {Wilson}},\ and\ \bibinfo {author} {\bibfnamefont
  {G.}~\bibnamefont {Rose}},\ }\bibfield  {title} {\bibinfo {title} {Quantum
  annealing with manufactured spins},\ }\href
  {https://doi.org/10.1038/nature10012} {\bibfield  {journal} {\bibinfo
  {journal} {Nature}\ }\textbf {\bibinfo {volume} {473}},\ \bibinfo {pages}
  {194} (\bibinfo {year} {2011})}\BibitemShut {NoStop}%
\bibitem [{\citenamefont {Georgescu}\ \emph {et~al.}(2014)\citenamefont
  {Georgescu}, \citenamefont {Ashhab},\ and\ \citenamefont
  {Nori}}]{Georgescu2014}%
  \BibitemOpen
  \bibfield  {author} {\bibinfo {author} {\bibfnamefont {I.~M.}\ \bibnamefont
  {Georgescu}}, \bibinfo {author} {\bibfnamefont {S.}~\bibnamefont {Ashhab}},\
  and\ \bibinfo {author} {\bibfnamefont {F.}~\bibnamefont {Nori}},\ }\bibfield
  {title} {\bibinfo {title} {Quantum simulation},\ }\href
  {https://doi.org/10.1103/RevModPhys.86.153} {\bibfield  {journal} {\bibinfo
  {journal} {Rev. Mod. Phys.}\ }\textbf {\bibinfo {volume} {86}},\ \bibinfo
  {pages} {153} (\bibinfo {year} {2014})}\BibitemShut {NoStop}%
\bibitem [{\citenamefont {Kandala}\ \emph {et~al.}(2017)\citenamefont
  {Kandala}, \citenamefont {Mezzacapo}, \citenamefont {Temme}, \citenamefont
  {Takita}, \citenamefont {Brink}, \citenamefont {Chow},\ and\ \citenamefont
  {Gambetta}}]{Kandala2017}%
  \BibitemOpen
  \bibfield  {author} {\bibinfo {author} {\bibfnamefont {A.}~\bibnamefont
  {Kandala}}, \bibinfo {author} {\bibfnamefont {A.}~\bibnamefont {Mezzacapo}},
  \bibinfo {author} {\bibfnamefont {K.}~\bibnamefont {Temme}}, \bibinfo
  {author} {\bibfnamefont {M.}~\bibnamefont {Takita}}, \bibinfo {author}
  {\bibfnamefont {M.}~\bibnamefont {Brink}}, \bibinfo {author} {\bibfnamefont
  {J.~M.}\ \bibnamefont {Chow}},\ and\ \bibinfo {author} {\bibfnamefont
  {J.~M.}\ \bibnamefont {Gambetta}},\ }\bibfield  {title} {\bibinfo {title}
  {Hardware-efficient variational quantum eigensolver for small molecules and
  quantum magnets},\ }\href {https://doi.org/10.1038/nature23879} {\bibfield
  {journal} {\bibinfo  {journal} {Nature}\ }\textbf {\bibinfo {volume} {549}},\
  \bibinfo {pages} {242} (\bibinfo {year} {2017})}\BibitemShut {NoStop}%
\bibitem [{\citenamefont {Monroe}\ \emph {et~al.}(2021)\citenamefont {Monroe},
  \citenamefont {Campbell}, \citenamefont {Duan}, \citenamefont {Gong},
  \citenamefont {Gorshkov}, \citenamefont {Hess}, \citenamefont {Islam},
  \citenamefont {Kim}, \citenamefont {Linke}, \citenamefont {Pagano},
  \citenamefont {Richerme}, \citenamefont {Senko},\ and\ \citenamefont
  {Yao}}]{Monroe2021}%
  \BibitemOpen
  \bibfield  {author} {\bibinfo {author} {\bibfnamefont {C.}~\bibnamefont
  {Monroe}}, \bibinfo {author} {\bibfnamefont {W.~C.}\ \bibnamefont
  {Campbell}}, \bibinfo {author} {\bibfnamefont {L.-M.}\ \bibnamefont {Duan}},
  \bibinfo {author} {\bibfnamefont {Z.-X.}\ \bibnamefont {Gong}}, \bibinfo
  {author} {\bibfnamefont {A.~V.}\ \bibnamefont {Gorshkov}}, \bibinfo {author}
  {\bibfnamefont {P.~W.}\ \bibnamefont {Hess}}, \bibinfo {author}
  {\bibfnamefont {R.}~\bibnamefont {Islam}}, \bibinfo {author} {\bibfnamefont
  {K.}~\bibnamefont {Kim}}, \bibinfo {author} {\bibfnamefont {N.~M.}\
  \bibnamefont {Linke}}, \bibinfo {author} {\bibfnamefont {G.}~\bibnamefont
  {Pagano}}, \bibinfo {author} {\bibfnamefont {P.}~\bibnamefont {Richerme}},
  \bibinfo {author} {\bibfnamefont {C.}~\bibnamefont {Senko}},\ and\ \bibinfo
  {author} {\bibfnamefont {N.~Y.}\ \bibnamefont {Yao}},\ }\bibfield  {title}
  {\bibinfo {title} {Programmable quantum simulations of spin systems with
  trapped ions},\ }\href {https://doi.org/10.1103/RevModPhys.93.025001}
  {\bibfield  {journal} {\bibinfo  {journal} {Rev. Mod. Phys.}\ }\textbf
  {\bibinfo {volume} {93}},\ \bibinfo {pages} {025001} (\bibinfo {year}
  {2021})}\BibitemShut {NoStop}%
\bibitem [{\citenamefont {Pfeuty}(1970)}]{Pfeuty1970}%
  \BibitemOpen
  \bibfield  {author} {\bibinfo {author} {\bibfnamefont {P.}~\bibnamefont
  {Pfeuty}},\ }\bibfield  {title} {\bibinfo {title} {The one-dimensional ising
  model with a transverse field},\ }\href
  {https://doi.org/https://doi.org/10.1016/0003-4916(70)90270-8} {\bibfield
  {journal} {\bibinfo  {journal} {Annals of Physics}\ }\textbf {\bibinfo
  {volume} {57}},\ \bibinfo {pages} {79} (\bibinfo {year} {1970})}\BibitemShut
  {NoStop}%
\bibitem [{\citenamefont {{Yang}}\ and\ \citenamefont
  {{Luo}}(2023{\natexlab{a}})}]{Yang2023a}%
  \BibitemOpen
  \bibfield  {author} {\bibinfo {author} {\bibfnamefont {Y.-T.}\ \bibnamefont
  {{Yang}}}\ and\ \bibinfo {author} {\bibfnamefont {H.-G.}\ \bibnamefont
  {{Luo}}},\ }\bibfield  {title} {\bibinfo {title} {{First-Order Excited-State
  Quantum Phase Transition in the Transverse Ising Model with a Longitudinal
  Field}},\ }\href@noop {} {\bibfield  {journal} {\bibinfo  {journal} {arXiv
  e-prints}\ ,\ \bibinfo {eid} {arXiv:2301.02066}} (\bibinfo {year}
  {2023}{\natexlab{a}})}\BibitemShut {NoStop}%
\bibitem [{\citenamefont {{Yang}}\ and\ \citenamefont
  {{Luo}}(2023{\natexlab{b}})}]{Yang2023c}%
  \BibitemOpen
  \bibfield  {author} {\bibinfo {author} {\bibfnamefont {Y.-T.}\ \bibnamefont
  {{Yang}}}\ and\ \bibinfo {author} {\bibfnamefont {H.-G.}\ \bibnamefont
  {{Luo}}},\ }\bibfield  {title} {\bibinfo {title} {{Pattern description of the
  ground state properties of the one-dimensional axial next-nearest-neighbor
  Ising model in a transverse field}},\ }\href
  {https://doi.org/10.48550/arXiv.2301.08891} {\bibfield  {journal} {\bibinfo
  {journal} {arXiv e-prints}\ ,\ \bibinfo {eid} {arXiv:2301.08891}} (\bibinfo
  {year} {2023}{\natexlab{b}})}\BibitemShut {NoStop}%
\bibitem [{\citenamefont {{Yang}}\ and\ \citenamefont
  {{Luo}}(2023{\natexlab{c}})}]{Yang2023d}%
  \BibitemOpen
  \bibfield  {author} {\bibinfo {author} {\bibfnamefont {Y.-T.}\ \bibnamefont
  {{Yang}}}\ and\ \bibinfo {author} {\bibfnamefont {H.-G.}\ \bibnamefont
  {{Luo}}},\ }\bibfield  {title} {\bibinfo {title} {{Topological or not? A
  unified pattern description in the one-dimensional anisotropic quantum XY
  model with a transverse field}},\ }\href
  {https://doi.org/10.48550/arXiv.2302.13866} {\bibfield  {journal} {\bibinfo
  {journal} {arXiv e-prints}\ ,\ \bibinfo {eid} {arXiv:2302.13866}} (\bibinfo
  {year} {2023}{\natexlab{c}})}\BibitemShut {NoStop}%
\bibitem [{\citenamefont {{Yang}}\ \emph {et~al.}(2023)\citenamefont {{Yang}},
  \citenamefont {{Chen}}, \citenamefont {{Cheng}},\ and\ \citenamefont
  {{Luo}}}]{Yang2023e}%
  \BibitemOpen
  \bibfield  {author} {\bibinfo {author} {\bibfnamefont {Y.-T.}\ \bibnamefont
  {{Yang}}}, \bibinfo {author} {\bibfnamefont {F.-Z.}\ \bibnamefont {{Chen}}},
  \bibinfo {author} {\bibfnamefont {C.}~\bibnamefont {{Cheng}}},\ and\ \bibinfo
  {author} {\bibfnamefont {H.-G.}\ \bibnamefont {{Luo}}},\ }\bibfield  {title}
  {\bibinfo {title} {{An explicit evolution from N\'eel to striped
  antiferromagnetic states in the spin-1/2 $J_{1}$-$J_{2}$ Heisenberg model on
  the square lattice}},\ }\href@noop {} {\bibfield  {journal} {\bibinfo
  {journal} {arXiv e-prints}\ ,\ \bibinfo {eid} {arXiv:2310.09174}} (\bibinfo
  {year} {2023})}\BibitemShut {NoStop}%
\bibitem [{\citenamefont {{Yang}}\ \emph {et~al.}(2024)\citenamefont {{Yang}},
  \citenamefont {{Chen}},\ and\ \citenamefont {{Luo}}}]{Yang2024a}%
  \BibitemOpen
  \bibfield  {author} {\bibinfo {author} {\bibfnamefont {Y.-T.}\ \bibnamefont
  {{Yang}}}, \bibinfo {author} {\bibfnamefont {F.-Z.}\ \bibnamefont {{Chen}}},\
  and\ \bibinfo {author} {\bibfnamefont {H.-G.}\ \bibnamefont {{Luo}}},\
  }\bibfield  {title} {\bibinfo {title} {{Identifying the ground state phases
  by spin-patterns in the Shastry-Sutherland model}},\ }\href@noop {}
  {\bibfield  {journal} {\bibinfo  {journal} {arXiv e-prints}\ ,\ \bibinfo
  {eid} {arXiv:2404.16330}} (\bibinfo {year} {2024})}\BibitemShut {NoStop}%
\end{thebibliography}

%

\end{document}